%% file: main.tex
\documentclass{sigchi}

\CopyrightYear{2020}
\setcopyright{acmlicensed}
\doi{https://doi.org/10.1145/3313831.3376235}
\isbn{978-1-4503-6708-0/20/04}
\conferenceinfo{CHI'20,}{April  25--30, 2020, Honolulu, HI, USA}
\acmPrice{\$15.00}

\usepackage{balance}       
\usepackage{graphics}      
\usepackage[T1]{fontenc}   
\usepackage{txfonts}
\usepackage{mathptmx}
\usepackage[pdflang={en-US},pdftex]{hyperref}
\usepackage{color}
\usepackage{booktabs}
\usepackage{tabu}
\usepackage{textcomp}

\usepackage{microtype}        
\usepackage{ccicons}          

\usepackage{todonotes}
\usepackage{wrapfig}
\usepackage{graphicx}
\usepackage{caption}
\usepackage{subcaption}
\usepackage[moderate,indent=normal, paragraphs=normal]{savetrees}

\def\plaintitle{}

\def\emptyauthor{}
\def\plainkeywords{Interactive systems; visualization; data science; contextual interviews; review analysis; text mining; opinion mining; sentiment analysis; schema generation.}

\makeatletter
\def\url@leostyle{%
  \@ifundefined{selectfont}{
    \def\UrlFont{\sf}
  }{
    \def\UrlFont{\small\bf\ttfamily}
  }}
\makeatother
\def\pprw{8.5in}
\def\pprh{11in}

\definecolor{linkColor}{RGB}{6,125,233}
\hypersetup{%
  pdftitle={\plaintitle},
  pdfauthor={\emptyauthor},
  pdfkeywords={\plainkeywords},
  pdfdisplaydoctitle=true, 
  bookmarksnumbered,
  pdfstartview={FitH},
  colorlinks,
  citecolor=black,
  filecolor=black,
  linkcolor=black,
  urlcolor=linkColor,
  breaklinks=true,
  hypertexnames=false
}

\DeclareRobustCommand{\subhead}[1]{\noindent\textbf{#1}} 
\DeclareRobustCommand{\subheadit}[1]{\noindent\textit{#1}}

\begin{document}

\title{Teddy: A System for Interactive Review Analysis}

\numberofauthors{6}
\author{
  \alignauthor{Xiong Zhang\\
    \affaddr{University of Rochester}\\
    \affaddr{Rochester, NY, USA}\\
    \email{xzhang92@cs.rochester.edu}}\\
  \alignauthor{Jonathan Engel\\
    \affaddr{Megagon Labs}\\
    \affaddr{Mountain View, CA, USA}\\
    \email{jonathan@megagon.ai}}\\
  \alignauthor{Sara Evensen\\
    \affaddr{Megagon Labs}\\
    \affaddr{Mountain View, CA, USA}\\
    \email{sara@megagon.ai}}\\
  \alignauthor{Yuliang Li\\
    \affaddr{Megagon Labs}\\
    \affaddr{Mountain View, CA, USA}\\
    \email{yuliang@megagon.ai}}\\
    \alignauthor{\c{C}a\u{g}atay Demiralp\\
    \affaddr{Megagon Labs}\\
    \affaddr{Mountain View, CA, USA}\\
    \email{cagatay@megagon.ai}}\\
  \alignauthor{Wang-Chiew Tan\\
    \affaddr{Megagon Labs}\\
    \affaddr{Mountain View, CA, USA}\\
    \email{wangchiew@megagon.ai}}\\
}

\maketitle
\input{abstract}


\printccsdesc
\input{intro}
\input{related}

\input{interview}

\input{results}

\input{criteria}

\input{teddy}

\input{evaluation}

\input{discussion}
\input{ack}
\balance{}

\bibliographystyle{SIGCHI-Reference-Format}
\bibliography{main}

\end{document}

%% file: abstract.tex
\begin{abstract}
Reviews are integral to e-commerce services and products. They contain a wealth of information about the opinions and experiences of users, which can help better understand consumer decisions and improve user experience with products and services.  Today, data scientists analyze reviews by developing rules and models to extract, aggregate, and understand information embedded in the review text. However, working with thousands of reviews, which are typically noisy incomplete text, can be daunting without proper tools. Here we first contribute results from an interview study that we conducted with fifteen data scientists who work with review text, providing insights into their practices and challenges.  Results suggest data scientists need interactive systems for many review analysis tasks. 
In response we introduce Teddy, an interactive system that enables data scientists to quickly obtain insights from reviews and improve their extraction and modeling pipelines.
\end{abstract}

%% file: intro.tex
\section{Introduction\label{sec:intro}}

Consumer reviews have become an essential part of e-commerce services and products, such as
hotels, restaurants, and job listings. Their prevalence is largely spurred by aggregator services such as booking sites for hotels, or restaurants, where reviews can help consumers decide between hotels or restaurants. Reviews are full of useful information, 
including consumer experiences, facts, tips and more. The abundance of reviews can 
provide reliable and relevant signals about the quality of services and products as well as how to improve them. Consumers regularly check reviews to inform their purchasing choices, online marketplace platforms display reviews along with summaries for consumers and sellers to facilitate their decision making. Business owners use reviews to track consumer feedback and adjust their products and services.  In a way, the collection of all user experiences with a product is the effective or true representation of the product. Therefore, consumer and enterprise services around products must leverage the distributional representations of the products embodied by reviews. Extracting insights from reviews can be widely useful to this end. 

Researchers across multiple fields, including data mining and natural language processing (NLP), have investigated the challenge of extracting, summarizing, and aggregating information from text and developed techniques for opinion mining and sentiment analysis~\cite{liu2012survey,pang2008opinion}.  Today, many e-commerce companies, especially those providing aggregation and search services, employ data scientists to analyze, extract, and summarize information from reviews. However, review text generated by consumers is notoriously noisy and often sparse in informational content. For example, a review about a hotel typically mentions only a couple of aspects about the hotel, such as cleanliness  and location, out of dozens of possible aspects. Reading and searching through thousands of sparse, noisy short texts in order to analyze, understand, and interpret them is a daunting task without effective tools. 

\begin{figure*}[!t]
\centering
\includegraphics[width=1\textwidth]{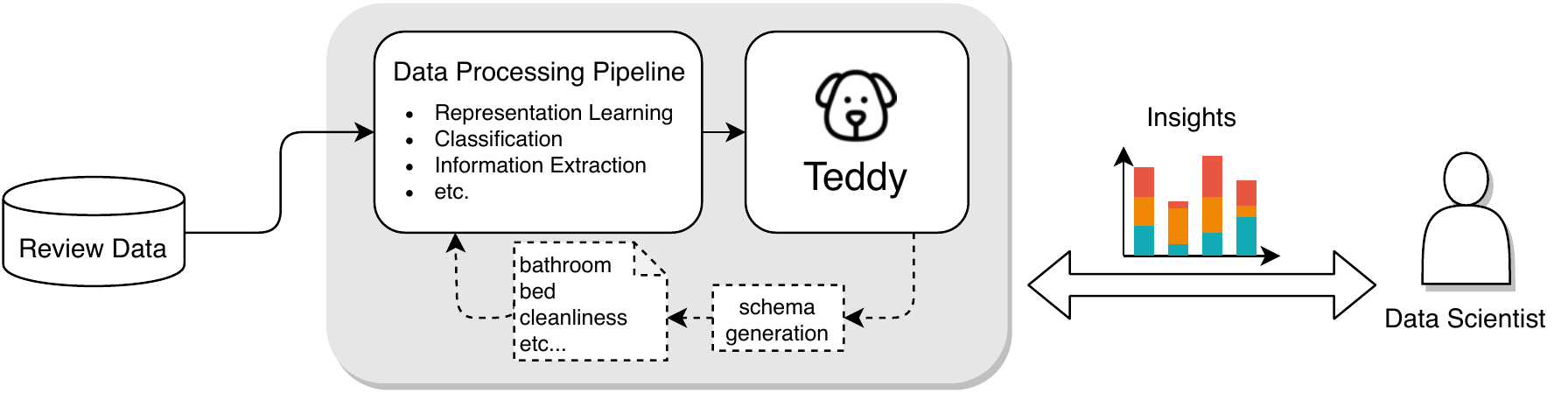}
\caption{The Teddy review exploration pipeline. Users can run the data processing pipeline based on their task, whether it is classification, opinion extraction, or representation learning, then use Teddy to gain insights about their data and model. They can also iterate on the data processing pipeline, for example by creating a new schema that describes attributes of their review corpus.\label{fig:data-prep-pipeline}}
\end{figure*}

In this paper, we first contribute results from an interview study that we conducted with fifteen participants to better understand the workflows and challenges of data scientists working on reviews. Our results suggest that data scientists spend most of their time in data preparation, a finding which provides additional evidence for similar findings from earlier general studies (e.g.,~\cite{kandel2012enterprise}). We find that data scientists are less concerned about developing new models or tuning hyper-parameters and are typically satisfied with using existing high-capacity language models such as BERT~\cite{devlin2019bert}. On the other hand, they are challenged by a lack of tools that would help across different stages of data preparation, ranging from labeling and crowdsourcing to interactive exploration and schema generation (where a schema is defined as a domain-specific set of attributes or aspects that users care about, for example a schema for the cell-phone domain might include price, weight, camera quality, etc.).  In particular, our findings suggest that data scientists need interactive tools to quickly obtain insights from reviews and to inform their extraction and modeling pipelines.

To address this need, we also contribute Teddy (Fig~\ref{fig:data-prep-pipeline}), an interactive visual analysis system for data scientists to quickly explore reviews at scale and iteratively refine extraction schemas of opinion mining pipelines. Informed by the results of our study, Teddy enables similarity-based multiscale exploration of reviews using fine-grained opinions extracted from them. Teddy is extensible, supporting common text search, filtering, regular expression matching, sorting operations, and their visualizations through programmatic extensions. Teddy sustains an interactive user experience for the analysis of large numbers of reviews by using a combination of pre-computation, indexing, and user-controlled separation of front- and back-end computations. Finally, Teddy enables data scientists to interactively revise and expand domain-specific schemas used for aspect-based opinion mining. We demonstrate the utility of Teddy through two in-depth use cases involving exploratory analysis of hotel reviews and iterative schema generation for opinion extraction from restaurant reviews, respectively.

We have made our research artifacts, including raw and aggregated data collected from the interview study and source code for Teddy, along with a running copy deployed as a Web application, available at~\url{https://github.com/megagonlabs/teddy}.

%% file: related.tex
\section{Related Work\label{sec:related}}
We build on earlier work in review text mining and interactive systems for visual review analysis. 

\subhead{Review Mining} Prior research applies sentiment analysis and opinion mining to extract and compile opinions and facts within large collections of review text~\cite{liu2012survey, pang2008opinion}. 
Sentiment analysis aims to quantify the sentiments expressed in text at different levels (sentence, paragraph, document, etc.). On the other hand, opinion mining builds on sentiment analysis to aggregate extracted sentiments into effective summaries to inform various tasks. Earlier research in sentiment analysis and opinion mining  proposes many approaches for mining the overall opinion at the document and sentence levels~\cite{kim2004determining,pang2002thumbs}.  Unsurprisingly, later research increasingly focuses on fine-grained extractions to derive opinions (including subjective judgments, facts, suggestions, and tips) per aspect or feature~\cite{hu2004mining,hu2004features,popescu2005extracting}. 
In the use cases presented here, we use OpineDB~\cite{li2019subjective} to extract opinions about domain-specific aspects (e.g., cleanliness, service, location, etc. in the hotel domain) from review text. The OpineDB extractor fine tunes BERT~\cite{devlin2019bert}, a pre-trained state-of-the-art language model, to perform fine-grained aspect-based extraction. Although Teddy benefits from OpineDB's high-quality interpretable extractions, it doesn't depend on it. Any other feature supervised or unsupervised extractor for text can be used at the data preparation stage, and our prototype includes an implementation of latent Dirichlet allocation (LDA)~\cite{blei2003latent} which can be selected as a feature extractor in a config file.  

\subhead{Visual Analysis of Reviews} Prior research introduces several tools for interactive review analysis and visualization. The general approach underlying these tools  is to first extract features from review text and then use interaction and visualization to facilitate the exploration of these attributes (e.g., sentiments, word frequency, etc.). Therefore, the differences among prior tools are in part characterized by the differences in their feature  extractions. 

Initial work on visual review analysis relies on coarse-grained (document or sentence level) sentiment extraction and visualizes  sentiments along with other textual features  using basic techniques, including scatterplot~\cite{morinaga2002mining}, rose plot~\cite{gregory2006user}, treemap~\cite{gamon2005pulse}, and graph~\cite{chen2006visual} visualizations.  Reviews contain richer information than document level or sentence level opinion visualizations can provide. Overall sentiments expressed in reviews can be sliced into finer-grained sentiments on domain-specific aspects.   
With the development of feature-based opinion mining,  researchers introduced feature-level opinion visualizations and tools that support coordinated views of these visualizations~\cite{alper2011opinionblocks,di2013sentiment,felix2016texttile, hao2013visual,liu2005opinion,oelke2009visual, soto2015exploratory,wu2010opinionseer,yatani2011review}. 

Note that our work here also falls into the general text visualization and visual text analytics  research,e.g.,~\cite{collins2009parallel,dou2013hierarchicaltopics,stasko2008jigsaw}.
A number of earlier work in text visualization focus on visual encoding design~\cite{collins2009docuburst, collins2009parallel,havre2000themeriver, wattenberg2008word}. Teddy addresses the  review text mining challenges informed by our interview study, the results of which prioritize the effective combination of visual analytic techniques over designing new visual encodings or interactions. Teddy is  akin to earlier general text analytics tools in its use of  clustering to avoid clutter for scalable topic exploration~\cite{dou2013hierarchicaltopics}, coupling text search with visualization ~\cite{stasko2008textsense}, and encoding textual similarity in two-dimensional layouts~\cite{chuang2012interpretation}.  We refer readers to existing surveys~\cite{kucher2018state,liu2018bridging} for a more complete discussion 
of the broader literature.  

Teddy supports a supervised fine-grained opinion extractor~\cite{liu2005opinion} as well as unsupervised topic modeling~\cite{blei2003latent} for feature-based review exploration.  Teddy takes multiple visualization techniques generally used in isolation by earlier approaches and combines them in a novel, information-rich configuration to enable the visual analysis of raw review text alongside fine-grained and aggregated opinion extractions and metadata. Unlike previous tools, which were typically designed for end users (e.g., customers), we designed Teddy for data scientists. It is extensible and supports common text operations needed by data scientists and their visualizations through programmatic extensions. In a novel approach, Teddy also facilitates an iterative improvement of opinion extractor schemas, extractions of which it already visualizes.

%% file: interview.tex
\section{Interview Study\label{sec:interview}}

To better understand data science practices and challenges in review analysis and mining, we conducted an interview study with data scientists working with review text corpora. 

\subhead{Participants}
We interviewed 15 researchers (12 male and 3 female) solicited from our professional networks. All participants were employed in technology companies. They worked at either AI/data-science research labs (10), review aggregating companies (2), hospitality-sector data processing companies (2) or job listing aggregator companies (1). Participants were geographically split between the United States (8), Japan (5), and Germany (2). 

Participants held job titles such as ``research scientist,'' ``data analyst,'' ``data scientist,'' ``research intern,''  or ``software engineer.'' One participant was a ``senior data scientist'', and one was a ``project manager.'' Henceforth, we will refer to all of the participants as ``data scientists.'' 
Most participants held PhDs (7 in computer science, 1 in industrial engineering). The rest were either interning PhD students (2) or held Master's degrees (2 in computer science, 2 in computational linguistics, 1 in applied math). 

\begin{figure}[!h]
    \centering
    \includegraphics[width=\linewidth]{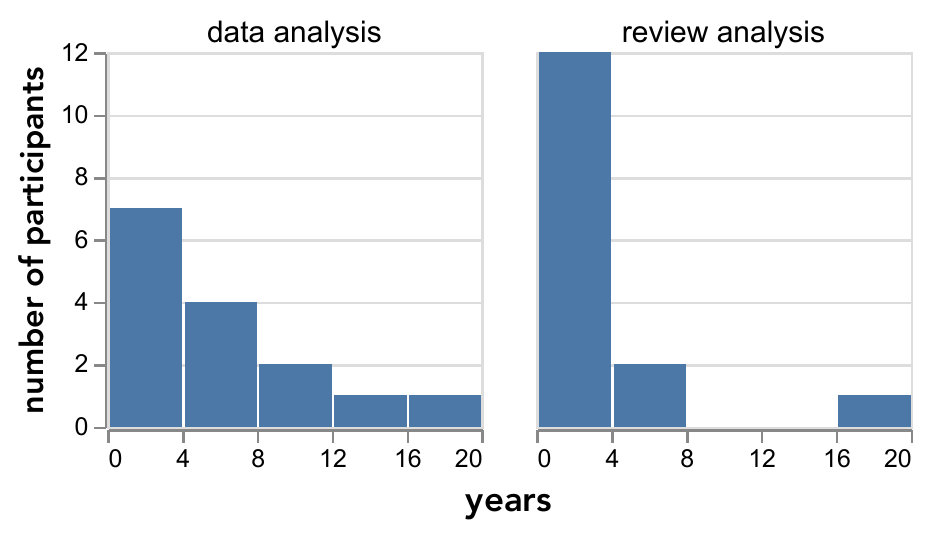}
    \caption{Experience of participants in data and review analysis. \label{fig:experience}}
\end{figure}

Past experience (Fig~\ref{fig:experience}) with review corpora ranged from less than 1 year to 18 years with a mean of 2.54 years (std=4.56).  Past experience with data analysis in general varied across participants, ranging from under 1 year to 20 years with a mean of 6.0 years (std=5.41). The inclusion of study participants with minimal experience in text data analysis allowed us to also incorporate the perspectives and challenges of new data scientists.

Most participants (10) worked with English-language corpora. The next most common language (4) was Japanese, and a final participant worked with Spanish and Portuguese corpora. Only 3 out of the 15 participants reported using multi-language corpora.

\subhead{Methodology}
Interviews were conducted in 1-hour sessions. With one exception, we interviewed one participant at a time, and with 1-3 researchers asking questions, taking notes, and recording audio. In the case of the exception, two subjects working on the same team were interviewed at the same time. For all interviews, written notes were compared and summarized between researchers. Once summaries were compiled, we sent them back to each corresponding participant and encouraged them to look over the summaries and reply with revisions or additional comments. Whenever possible, we interviewed participants in person, otherwise resorting to video conferencing.

We asked participants open-ended questions about their work experiences with review corpora, and asked them to walk us through specific examples of projects that they have recently worked on. We designed a rubric of thirteen questions that would identify key elements of data scientists' workflows. The rubric included broad questions such as \textit{``How do you prepare your data for use in your pipeline?} and \textit{``What do you spend most of your time on?''}, and more specific questions such as \textit{``What data sources and formats do you use?''} We also asked participants additional questions that arose from their other answers. At the end of each interview, we asked participants for additional comments if they felt that we had missed important information about their experiences working with reviews.

Once interviews were completed, we used an iterative coding method to analyze the notes. One of our interviewers summarized the interviews from independently taken notes, and another interviewer coded the summarized interviews using an inductive iterative coding method while receiving feedback from the first interviewer. Common experiences and tasks were collected into groups, and we refined these categories as further data was analyzed.

%% file: results.tex
\subsection{Results}

\subhead{Task Taxonomy}
We identified 3 overarching task types that participants described in their interviews.

{\em Classification}, where analysts develop algorithms to classify either entire reviews or individual sentences into predefined categories (e.g., based on sentiment or between a set of topics).

{\em Extraction}, where analysts develop algorithms to detect relevant entities from reviews, as well as descriptive text elsewhere in the reviews that directly modify or describe the entities (e.g., extracting opinions about the quality of specific types of food from restaurant reviews). 

{\em Representation}, where analysts build graph representations or database architectures to accommodate data and insights related to review text corpora (e.g., developing a schema of amenities offered by specific hotels).

The distribution of participants between these categories of work was roughly even: entity extraction and classification tasks were the primary work categories for 4 participants each, entity representation was the primary work category for 5 participants. One participant described a work pipeline that included equal amounts of entity representation and extraction tasks, and a final participant described their experiences across a variety of relevant projects comprising entity representation and classification tasks. 

\subhead{Data}
 Participants generally sourced data directly from the client companies commissioning the technologies that the participants were developing. For participants engaged in more open-ended research, public domain data was used. In several cases, participants used both proprietary and public domain data in their work. Almost all participants reported needing to look at raw data points (i.e., review text)  frequently as part of their analysis. Roughly half of the participants reported that their review analysis tasks required data cleaning 
 as a preliminary task.

Dataset sizes varied across participants, ranging from sets with thousands of data points to sets with tens of millions of data points. Participants using datasets with fewer than 1 million examples generally used CSV files to store their data; those working with larger datasets used SQL databases. Data scalability was named as a significant concern for pipeline management for 5 out of the 15 participants. Counter-intuitively, those working on the largest datasets with tens of millions of entries did not report any concerns about the scalability of their pipelines (3/15 participants). 

We believe that this discrepancy is due to the nature of the tasks being performed by the analysts: all three participants that reported having no scalability concerns despite using datasets with tens of millions of entries were working on entity representation tasks that did not require the training of computationally expensive machine learning models. These findings suggest that there is a need for future work that rigorously analyzes the different needs of data scientists working on different tasks.

\begin{figure}[!t]
    \centering
    \includegraphics[width=0.75\linewidth]{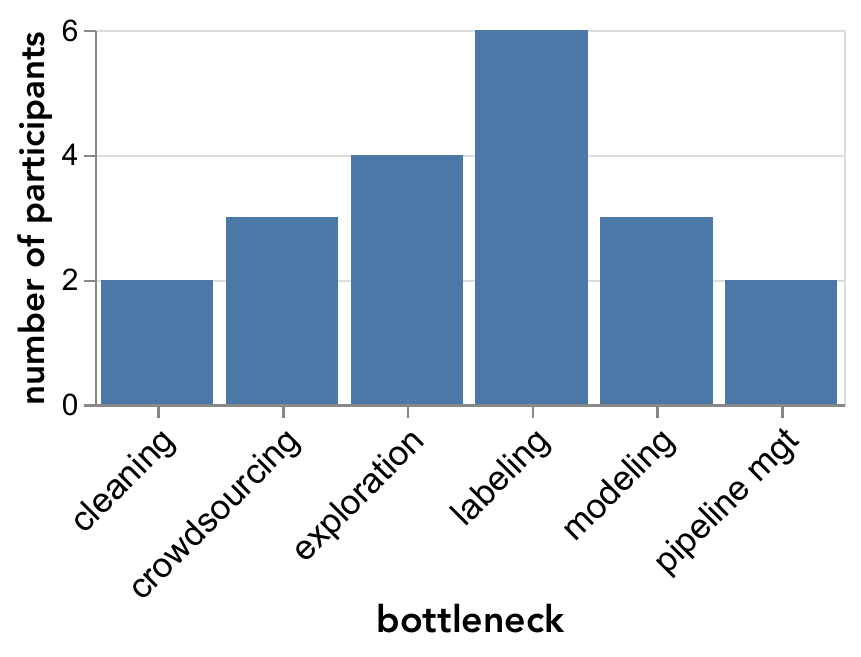}
    \caption{We identify some of the most common bottlenecks reported by data scientists working with review data. We find most bottlenecks involve labeling and exploration.}
    \label{fig:aggregated-bottlenecks}
\end{figure}

\subhead{Tools}
Python was ubiquitous as the primary programming language for our participants, with some also using Java, and SQL was used for database interfacing. This finding is consistent with a trend reported by a previous interview study of data scientists~\cite{Demiralp:2017:DSIA}. Participants generally used a combination of Gensim~\cite{rehurek_lrec}, NLTK~\cite{BirdKleinLoper09}, and SpaCy~\cite{spacy2019} for natural language processing (NLP) functions and word embeddings. One participant specifically mentioned that their team was transitioning from NLTK to SpaCy, as they found SpaCy to outperform NLTK in all of their use-cases: \textit{``The quality and performance [of tokenization and other NLP functions] are much better on SpaCy.''}

Participants working with Japanese corpora used SudachiPy~\cite{TAKAOKA18.8884} for their core NLP work. Participants working directly with neural network models used BERT~\cite{devlin2019bert} as their model framework. Roughly one quarter of participants (4) reported using sentiment extraction as part of their analysis, using NLTK, SpaCy, or self-developed algorithms to perform this extraction.  Four of the participants reported using Jupyter Notebook~\cite{jupyter} as a collaboration tool.


\subhead{Model Training}
While most participants (12) had some sort of model training as part of their pipeline, most were interested in applying off-the-shelf models rather than designing new ones. Only two participants, working on the same project, specified parameter tuning as part of their work pipeline, and only three participants (including the aforementioned two) reported having several different options for which models to use for their pipeline. By contrast, half of the participants reported that they regularly conducted manual reviews of individual data points as part of a model debugging process.

When asked to specify which metrics were their top priorities for improvement when optimizing their models, participants were equally likely to report either precision or recall as their top priority. One participant said that they were focusing on finding higher quality trainig data, rather than trying to improve their model's performance through hyperparameter tuning or changing the model architecture.

\begin{figure}[!t]
    \centering
    \includegraphics[width=\linewidth]{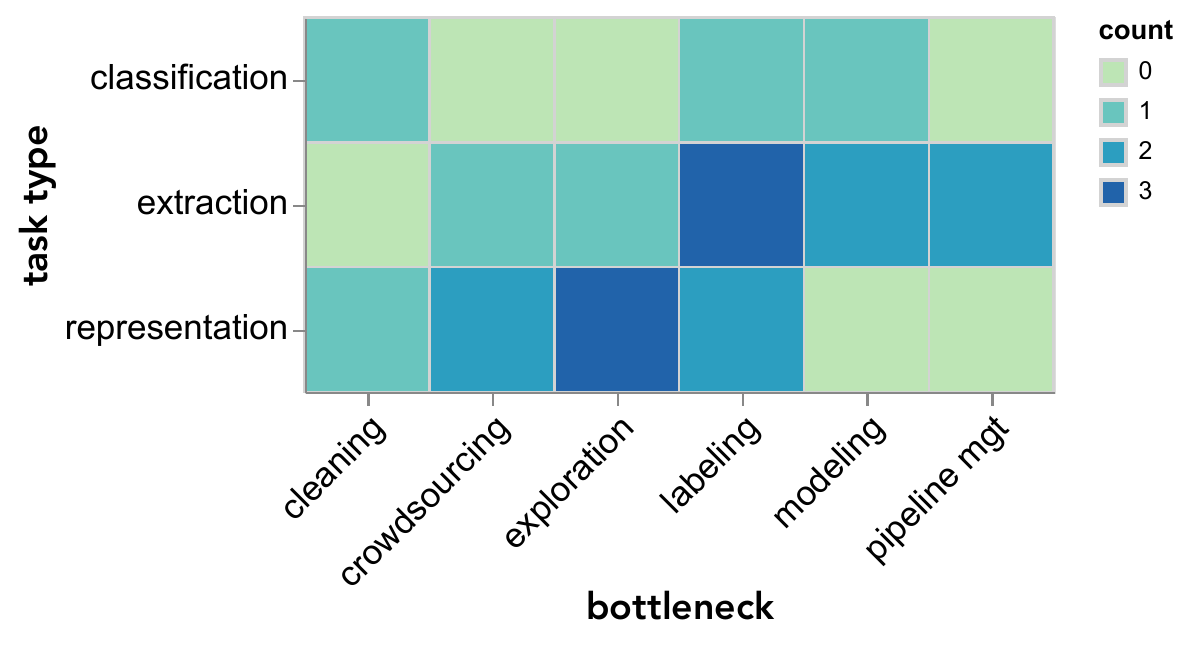}
    \caption{Bottlenecks reported by data scientists working with review data, separated by the type of task. Most bottlenecks involve labeling and exploration, especially among those working on entity extraction and representation.}
    \label{fig:interview-results-1}
\end{figure}

\subhead{Challenges} Participants reported a variety of different tasks as being particularly challenging 
bottlenecks to their work (Figs~\ref{fig:aggregated-bottlenecks} and  \ref{fig:interview-results-1}), and expressed desires for various different types of software tools which could help them overcome these difficulties. 

\subheadit{Data Cleaning} One quarter of the participants who included data cleaning as part of their pipeline (2/8) described data cleaning as a bottleneck. However, only one participant specifically requested additional data cleaning tools. For our purposes, we define ``data cleaning'' as tasks that involve editing or removing data at the outset of the pipeline, such as removing reviews that are too short,  censoring foul language,  or applying spellchecking functions to the data. Difficulties with data cleaning often stemmed from the scale of the data to be cleaned, and the necessity of subjective judgments as part of the cleaning process, making automation difficult or impossible: \textit{``you have this problem which is data sparsity. Consider suggestion mining, out of millions of examples, only one or two percent of them are suggestions. It is very inefficient. Our goal in cleaning is to make the dataset more focused and ignore the negative sentences.''} We also found that data scientists with more seniority were less likely to identify data cleaning as a bottleneck than more junior data scientists. A possible explanation for this phenomenon is that data cleaning tasks are more likely to be delegated to less experienced team members.

\subheadit{Data Labeling}
Many participants (6/15) identified
data labeling as a bottleneck across all task categories, and expressed a desire for data labeling tools.
Similar to our findings for data cleaning, data labeling being reported as a bottleneck correlated negatively with years of experience. As before, we think that the delegation of labeling tasks to less experienced team members is a likely explanation for this phenomenon.

\subheadit{Crowdsourcing}
For those utilizing crowdsourcing (9/15), one third reported that crowdsourcing was a bottleneck, due to the difficulties of designing proper crowdsourcing tasks and verifying that crowdsourced labels were accurate. Participants working on entity representation were the most likely to identify crowdsourcing as a bottleneck.

\subheadit{Data Exploration}
Four participants, working on either entity extraction or entity representation tasks reported that initial data exploration was a significant bottleneck, and expressed a desire for better tools for data exploration. One participant identified the particularly high stakes associated with this task, which further slowed down its implementation: \textit{``once we agree on a schema with the client we can't change it anymore, so we need to be really careful to get it right the first time.''}

Three participants, working on either entity extraction or entity representation tasks, requested integrated search functions that would allow them to search for specific lexical features in their dataset, as opposed to simple string-matching. 
Two participants also expressed a need for a search function that would allow them to find reviews similar to a selected review.

 Data visualization tools were also a commonly reported desired capability. This need was even more frequently expressed by participants working on entity representation tasks.

\subheadit{Modeling}
Two participants working on entity extraction tasks and one working on classification tasks identified model training as a bottleneck to their workflow, expressing frustration with slow runtimes for training algorithms. Generally, participants did not seem to consider model training to be a significant bottleneck to their work, even when asked about it specifically. One participant said \textit{``once you start the model, it takes one hour, two hours, and then you're done.''}

 Even when working to debug machine learning models, participants seemed more focused on fine-tuning training datasets than on changing model architectures or hyperparameters for training algorithms. These responses suggest to us that while data scientists will certainly still benefit from advances in the state of the art of machine learning algorithms, there is also a need for tools that better facilitate the collection, preparation, and exploration of data. 
 
 We found a correlation between years of experience and the likelihood of reporting model training as a bottleneck. It is possible that this phenomenon could also be explained by the division of labor hypothesis proposed earlier.

\subheadit{Pipeline Management}
Two participants, both working in entity extraction tasks, identified bottlenecks in managing the various parts of their pipeline. Despite this low proportion of participants identifying pipeline management as a bottleneck, two fifths (6/15) of the participants nevertheless expressed a desire for better pipeline management tools. Participants reported that they found it tedious to switch between tasks, which would often require them to go from using command line interfaces to writing scripts or examining data files manually and back again. They also expressed frustration with having to manage multiple overlapping data files and losing track of how the files corresponded to each other. 

\subsection{Generalizability}
Our study focused on a specific domain with low workflow variation, targeting a more homogeneous group than earlier broader studies,e.g,~\cite{Demiralp:2017:DSIA,kandel2012enterprise,kandogan2014data,muller2019data}. We also note that research on qualitative studies suggests that saturation occurs with twelve participants on average for homogeneous groups~\cite{guest2006}, and that sample sizes under twenty are preferable for the solicitation of fine-grained information~\cite{crouch2006}, findings in line with our experience conducting this study. 

Participants in the study were recruited from our professional network, forming a partial representation of the broader community of data scientists working on reviews. Our findings are nonetheless valuable, as our study is primarily intended to inform the development of systems that can address open problems in review text analysis, rather than providing a comprehensive survey of the field as a whole.

Review text analysis shares characteristics with text analysis at large. While further studies are warranted, we believe our findings can also inform the broader tooling research and practice for text analysis, particularly short text analysis.

%% file: criteria.tex
\section{Design Considerations\label{sec:criteria}} 

Informed by the interview study results above and our own experience along with a prior data exploration paradigm~\cite{Shneiderman:1996:ETD}, we derived 5 design considerations for the Teddy prototyping system.

\subhead{D1} Users should be able to explore, inspect, and compare clusters of review text based on the semantic similarity as determined by the opinions and topics within them. 

\subhead{D2} Users should be able to explore domain-specific aspects of reviews, then revise and build on existing extractor schemas in an iterative manner.

\subhead{D3} Users should be able to programmatically control the data displayed. In particular, the prototype should support sorting, filtering, and pattern matching on raw text and extracted features.

\subhead{D4} Users should be able to get an overview of the reviews from all the entities or from a single one.

\subhead{D5} Users should be able to zoom in and out the data visually and semantically in a multi-scale fashion, accessing details for entities and reviews on demand.


%% file: teddy.tex
\begin{figure*}[tbh]
    \includegraphics[width=\textwidth]{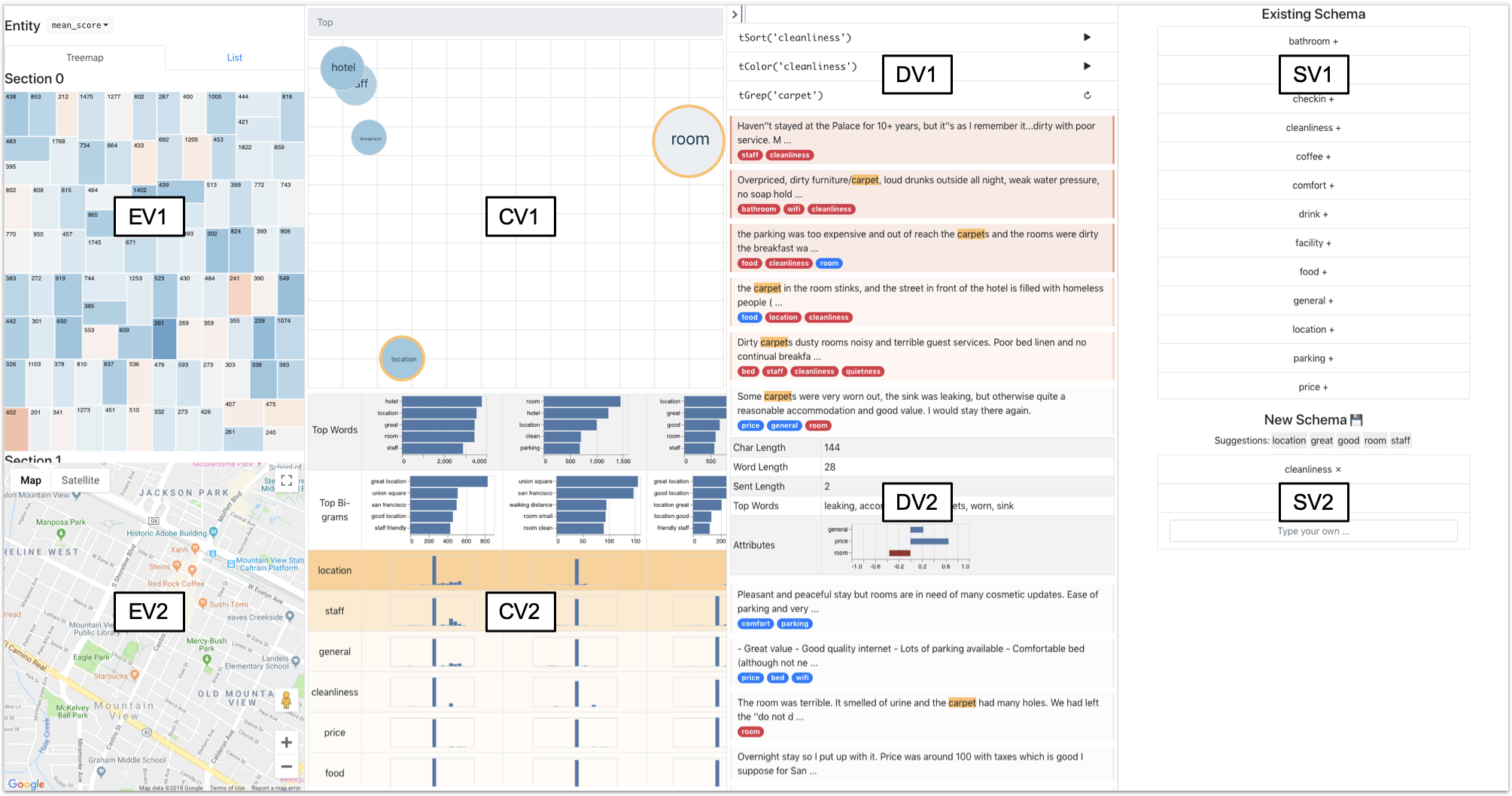}
    \caption{Overview of the Teddy user interface. From left to right: \textbf{Entity View (EV)} displaying the entities with Treemap~/~List (EV1) and an optional map (EV2); \textbf{Cluster View (CV)} displaying hierarchical clusters (CV1) and aggregated cluster statistics (CV2); \textbf{Detail View (DV)} displaying individual reviews items (DV2) and command line interface to filter/sort reviews; \textbf{Schema Generation View (SV)} for reviewing the existing schema (SV1) and building new ones (SV2). Details about these ``Views'' are described in the Teddy System Design section. \label{fig:teddy-overview}}
\end{figure*}

\section{Teddy System Design}\label{sec:teddy}
With the proposed design considerations in mind, we developed Teddy\footnote{Teddy = \textbf{T}ext \textbf{e}xplorer for diving into \textbf{d}ata \textbf{d}eepl\textbf{y}} an interactive system for review analysis. Teddy runs as a single-page web app, along with a backend server that stores data and provides various functionalities for text analytics. The backend server is implemented with Flask~\cite{flask}, Pandas~\cite{pandas}, and scikit-learn~\cite{scikitlearn}, while the frontend is implemented with D3~\cite{d3}, Vega-Lite~\cite{vegalite}, and React~\cite{react}. 

Teddy is designed to work with review datasets from any domain where reviews are associated with specific entities (e.g., hotels, restaurants, companies).

\subsection{Data Processing Pipeline}

In order to generate cluster visualizations and useful summary statistics for the text data, we implemented a preprocessing pipeline. There are three major parts in our preprocessing pipeline:

\subhead{Feature Vectorization} The first step is to generate feature vectors from the review text. In order to be adaptable to different datasets, Teddy provides users with a configuration file to specify a featurization algorithm. Currently, Teddy supports the use of Opine~\cite{li2019subjective} and latent Dirichlet allocation (LDA)~\cite{blei2003latent}. If Opine is selected, the user needs to provide a flat schema comprising a list of strings corresponding to topics mentioned in the dataset, whereas LDA does not use a predefined schema. The feature vectors generated in this step are N-dimensional, where N is a parameter specified in the configuration file. In the figures and use cases described in this paper we used a setup of 21-dimensional attribute vectors extracted by Opine.

\subhead{Clustering} Next, we run a series of K-means clustering algorithms to generate nested clusters over the feature vectors. Feature vectors are  clustered into $K1$ groups. Then, for each cluster, vectors assigned to it are further clustered into another $K2$ groups of subclusters. This process repeats until either a recursion depth of $d$ is reached or until there are insufficient data points in a given cluster for further clustering to be performed. Parameters for clustering can also be specified in the configuration file. In our figures, we used a setup of $K1=5$, $K2=3$, $d=5$, which we found to be suitable parameters for exploring our dataset.

\subhead{Summarization} Finally, for each cluster, we compute the following items based on the reviews in the cluster: the average number of characters, words, and sentences; top-N words and bi-grams (calculated using TFIDF); histogram distributions of scores for each schema attribute, and the averaged attribute scores. When using LDA, aggregate sentiment scores are also calculated in this step, whereas in Opine this is calculated as part of feature extraction. These summaries are displayed in the app in order to allow users to more easily see differences between clusters.


\subsection{Data Exploration in Views}

Teddy's interface has four sections or ``Views.'' These views, shown in Fig.~\ref{fig:teddy-overview}, are outlined below.

 \begin{figure}
    \includegraphics[width=0.48\textwidth]{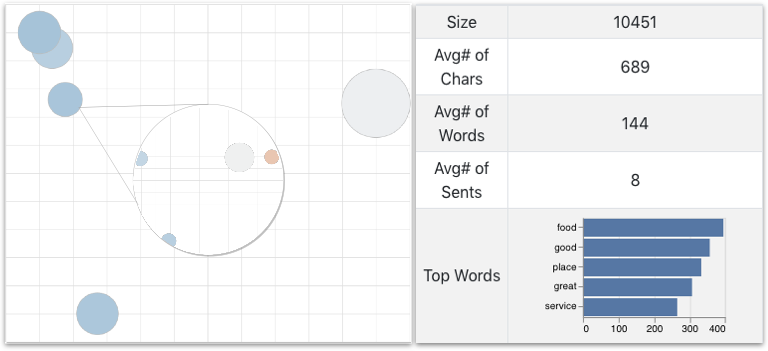}
    \caption{The Cluster View provides a scalable way to explore aggregate statistics of reviews. Double-clicking on a cluster shows a new clustering over the selected subset of reviews (shown here as an inset). The color represents the average sentiment of reviews in the cluster, while the size represents the number of reviews. When a cluster is selected, data such as frequently occurring words and histograms over the sentiment for each aspect are displayed (example on the right), and the user can select a second cluster to compare these statistics (shown in Fig.~\ref{fig:rosa-select-2}).}
    \label{fig:cluster_zoom}
\end{figure}
    
 \subhead{Entity View} shows the entities and allows users to select and load entity-specific clusters. Users can toggle between two kinds of visualizations with the tabs on the top (Fig.~\ref{fig:teddy-overview}~-~EV1). Treemap visualization groups similar entities and allocates different sizes of blocks based on the amount of reviews associated with the entity, while the list visualization allows the user to clearly see entity names. Each entity item is colored by an averaged attribute score selected from a drop-down menu. When the user clicks on an entity, the map on the bottom displays its location if coordinates are available in the dataset (Fig.~\ref{fig:teddy-overview}~-~EV2). An information card also pops up with a preview picture of the entity, the entity address (if available), and a button to load hierarchical clusters of all the reviews for that entity. 
In the absence of entity information, Teddy disables the treemap and map views, but the rest of the application stays functional.
   
\begin{figure}[tbh]
    \includegraphics[width=0.48\textwidth]{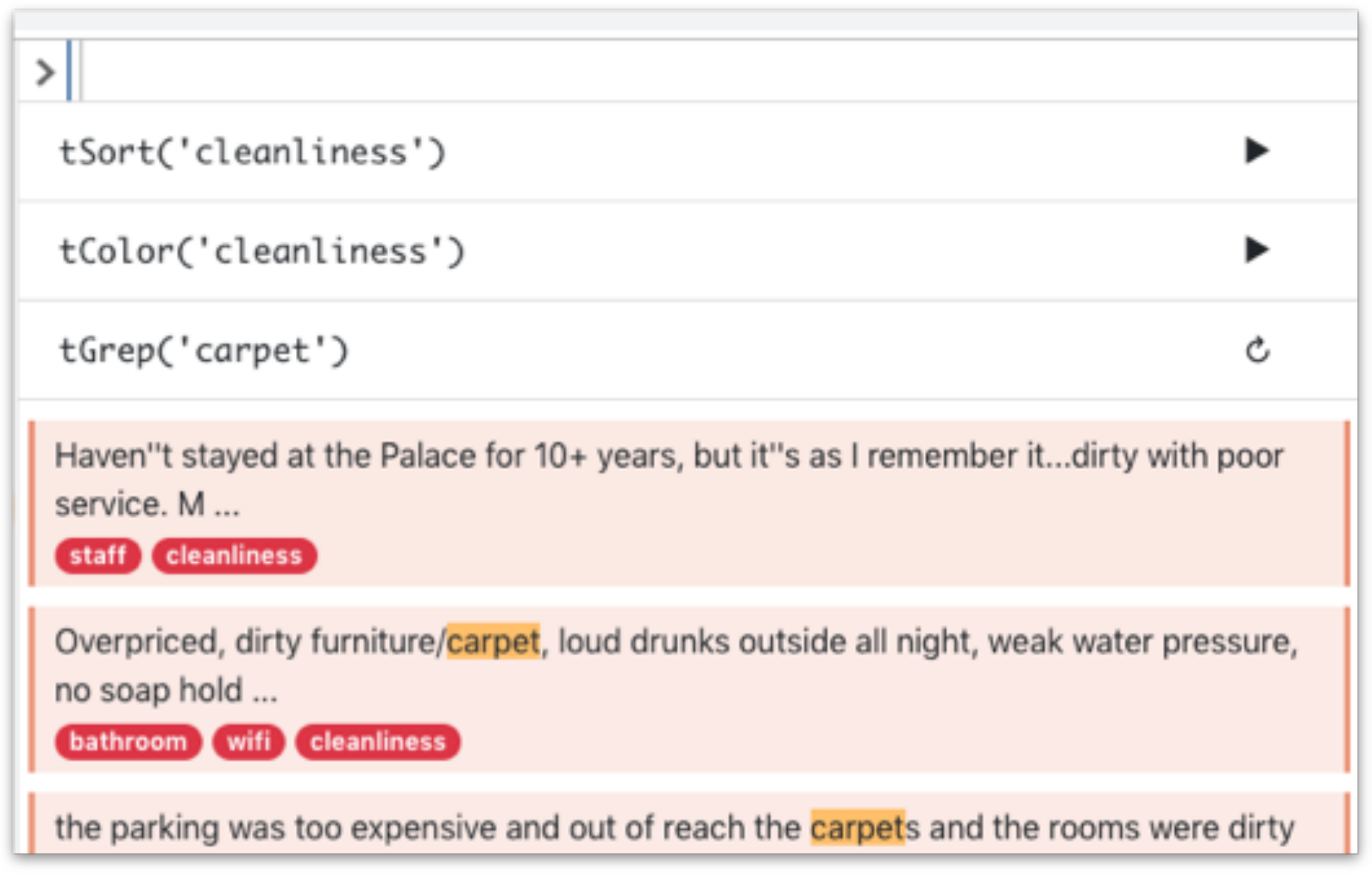}
    \caption{Teddy command line interface for operations on reviews (outlined in Table~\ref{tab:commands}). In this example we explore the relationships between cleanliness and carpet quality at hotels. We filter reviews with the keyword ``carpet,'' then sort and color the reviews based on the cleanliness attribute, revealing that people write about the carpet in mostly negative contexts in our dataset.}
    \label{fig:command-line}
\end{figure}

\subhead{Cluster View} enables the efficient navigation of the dataset by showing clusters of similar reviews and their statistical summaries. Users can explore the cluster hierarchy by zooming in and out, or compare different clusters to get insights (Design considerations D1 and D5). This view also improves the perceptual scalability of the system. For a description of how the clusters are generated, please see the Data Processing Pipeline section of this paper.

In the Cluster View, Teddy initially shows clusters for the entire dataset for an overview (D4). These clusters are plotted as colored circles in a 2D space, based on PCA projections of the cluster's centroids (Fig.~\ref{fig:teddy-overview}~-~CV1). The radius of the circle corresponds to the size of the cluster, the color (from red to blue) corresponds to the averaged sentiment score of the review texts in the cluster. Clusters are labeled with the name of the schema topic that made the largest contribution to its sentiment score in the direction of its overall sentiment valence. To avoid confusion, if two clusters have their largest sentiment score contributions from the same topic and have overall sentiment scores with the same valence, the cluster with the larger contribution will take the label and the other cluster will use its second-most-contributing topic as its label.

When the user clicks on a cluster circle, Teddy shows the summarizing information in the bottom table (Fig.~\ref{fig:teddy-overview}~-~CV2). This includes the pre-computed summarizations: average number of characters, words, and sentences; top-N words and bi-grams (based on TF-IDF); histogram distributions of each attribute; and the averaged attribute scores. The table also shows the same set of statistics for the whole dataset for context. A user can compare two clusters side-by-side in the table by cmd-clicking (shown in Fig.~\ref{fig:rosa-select-2}). Teddy will calculate the histogram distances between the clusters for each attribute, and highlight the ones with the largest difference. This indicator can help users to locate interesting parts to keep an eye on later when they see the raw review text. To zoom into the next level of clusters (show the clusters inside a cluster), users can simply double click on the circle (Fig.~\ref{fig:cluster_zoom}). To zoom out, users can click on the directory style links on the top (D5).

\subhead{Detail View} shows raw review texts sampled from whichever cluster(s) the user has clicked on. To avoid flooding the screen, Teddy requests 10 reviews at a time, and a ``Load More'' button is available to request 10 more review items. The reviews are truncated into 100 characters to improve readability, but a click on each of them will show the full text and a panel with summarizing information similar to the clusters (Fig.~\ref{fig:teddy-overview}~-~DV2), which is also pre-computed as part of the preprocessing pipeline. Reviews are also labeled with any schema attributes that were detected in the review, colored according to their attendant sentiment score. This allows the user to quickly see important information about the review without reading the entire text.

\begin{table}[!b]
\small
\centering
\begin{tabu}{cXc}
\toprule
Command & Functionality & Remote\\
\midrule
 tSort & \textit{Sort reviews based on an attribute} & Y \\
 tFilter & \textit{Filter reviews based on an attribute and a predicate function} & Y \\
 tGrep & \textit{Search for reviews which texts have a certain pattern} & Y \\
 tColor & \textit{Change the background color of review items based on a given attribute} & N \\
 tReset & \textit{Reset previous operations} & Y \\
\bottomrule
\end{tabu}
\caption{Teddy commands for customizing the reviews shown. The ``Remote'' column corresponds to whether the command can also be run offline by the server on all the reviews. 
\label{tab:commands}}
\end{table}

In the interview study, we found that many participants used bash scripts to search and filter their datasets. To emulate a similar experience, in the Detail View users can also sort or filter the reviews based on an attribute or a specific pattern. Teddy provides a set of commands (shown in Table \ref{tab:commands}) to programmatically query reviews. For simplicity, the commands are JavaScript functions. Users can type commands in the input prompt (Fig.~\ref{fig:teddy-overview}~-~DV1) at the top, then hit ``Command+Enter'' to run. Teddy first runs the command on the current reviews that are displayed on the frontend. This helps the user to debug and tweak their command quickly. All the functions are performed on either the text level (grep) or the feature level (sort, filter, color). When the user is satisfied with the command, they can use the ``Remote Run'' button on the right to evaluate the command on the server end and get results from the whole dataset. In this step, the JavaScript-based code will be parsed into command names and arguments, then sent to the backend to be evaluated in the server environment. This separation provides a two-step lazy-evaluation on the commands, thus improving Teddy's computational scalability. Every new command will be performed on the results of the previous one, so the user can gradually build a set of operations to re-produce the discovery process (D3).

\subhead{Schema Generation View} is a workspace for building a new schema. Since the schema is assumed to be flat, it is shown as a list of words. Whenever users identify new attributes mentioned in review text, they can add them to the ``New Schema'' section (Fig.~\ref{fig:teddy-overview}~-~SV2), in addition to importing attributes from the existing schema (Fig.~\ref{fig:teddy-overview}~-~SV1). To aid this discovery process, Teddy displays the most frequent words in the selected cluster(s) as suggestions for the new schema. Users can edit the new schema and export it to be used as an input to re-run the data preparation step. By repeating this process, users can refine the schema to better understand the dataset (D2).

%% file: evaluation.tex
 \section{Usage Examples}
As an evaluation of Teddy, we observed two research software engineers, Rosa and Chao, in our lab apply the system for completing various review analysis tasks. Both Rosa and Chao work with review text regularly and neither of them was a participant in our interview study.  
 
\subsection{Exploratory Analysis of Hotel Reviews}
\begin{figure}[t!]
    \centering
    \includegraphics[width=\linewidth]{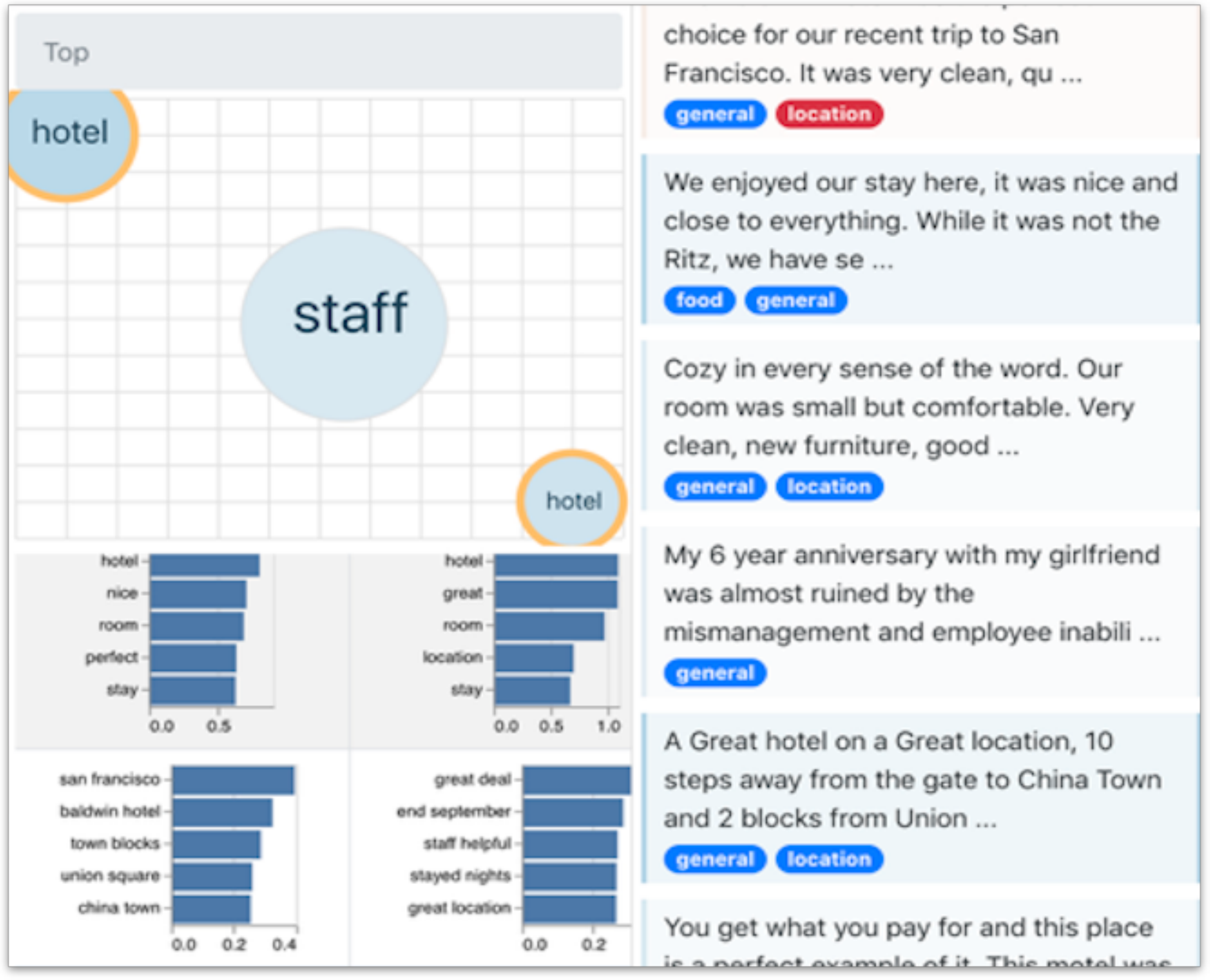}
    \caption{Two clusters have been selected from the visualization in the top left corner. In the bottom left, we can see histograms corresponding to the top-5 words and top-5 bigrams for each selected cluster. Along the right, we see excerpts of reviews tagged with their attribute category bubbles colored according to their sentiment scores.}
    \label{fig:rosa-select-2}
\end{figure}


We tasked Rosa with analyzing the customer reviews of a set of hotels to identify specific areas in which the hotels were either exceeding expectations or under-performing, and to extract quotes from reviews that exemplify the hotels' strengths and weaknesses. Additionally, we asked her to help debug an attribute extractor that we are developing, and which is being used in conjunction with Teddy, and to provide suggestions for a new attribute schema so that other researchers on her team can develop more fine-grained quality scores for her dataset.
 
Upon opening the app, Rosa saw that the data has been sorted into three clusters, two of which were labeled ``hotel'' and one of which was labeled ``staff.'' Rosa wanted to learn more about the differences between the two ``hotel'' clusters, so she command-clicked on them to get a side by side comparison (Fig~\ref{fig:rosa-select-2}). While the ``top words'' histograms was similar, the ``top bigrams'' histograms showed a significant difference: for cluster 1, 4 out of the top 5 bigrams were about the hotels' locations, whereas for cluster 2 only 1 out of the top 5 bigrams had to do with hotel locations. Rosa wanted to find information about the hotels' rooms, so she zoomed in on cluster 2 by double-clicking on it.

\begin{figure}
 \begin{subfigure}{.23\textwidth}
     \includegraphics[width=\textwidth]{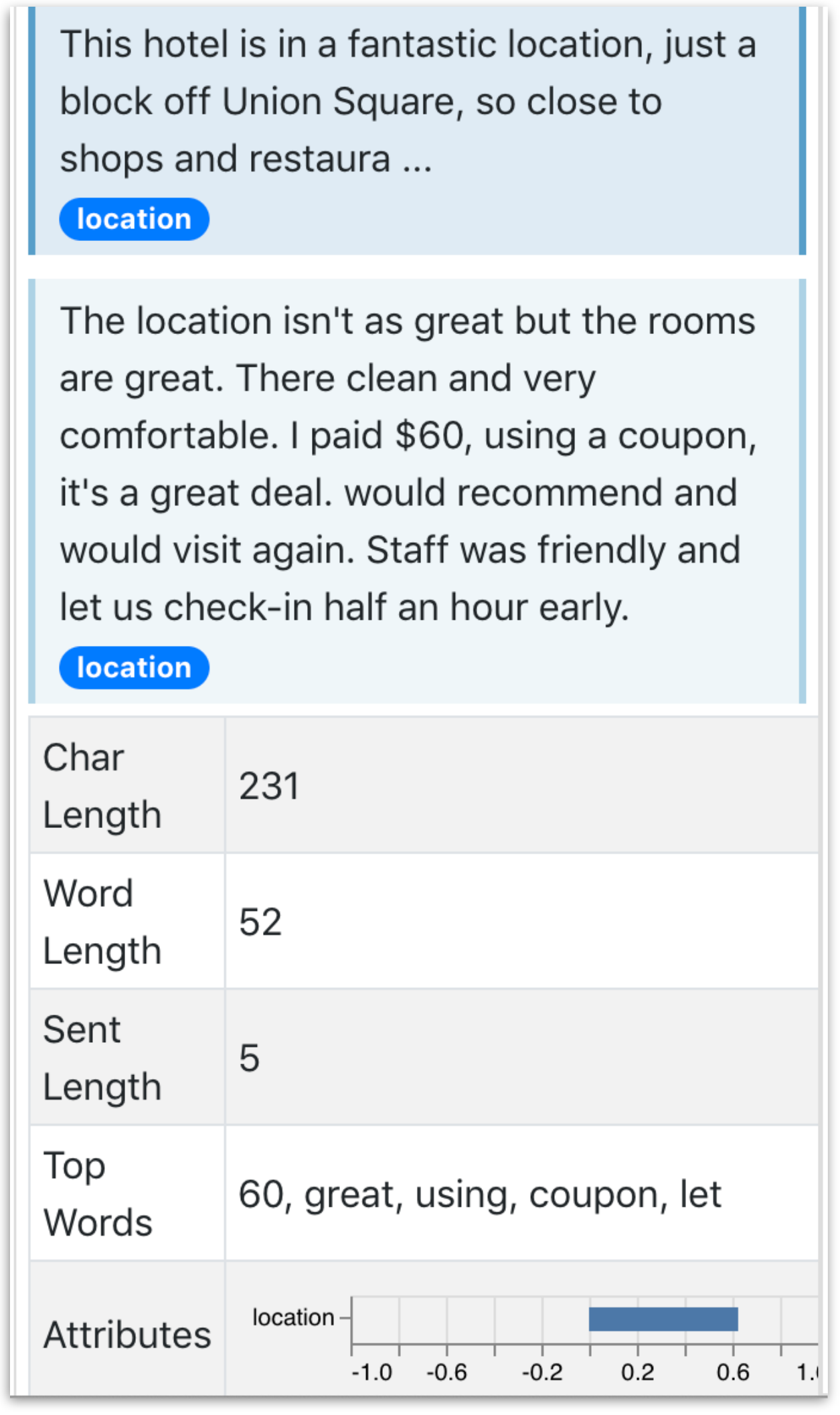}
     \caption{\label{fig:rosa-review-1}}
 \end{subfigure}
 \begin{subfigure}{.23\textwidth}
        \includegraphics[width=\textwidth]{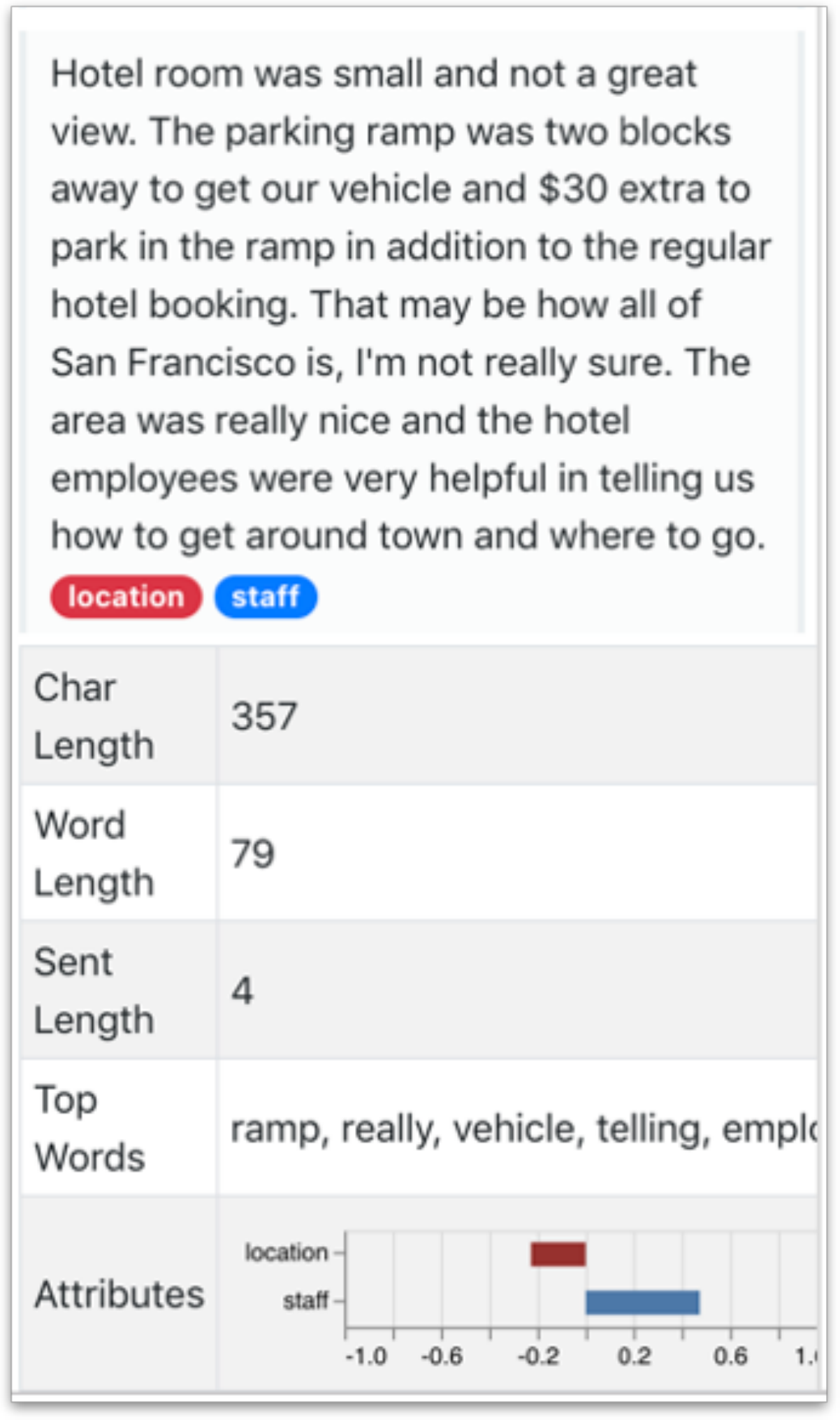}
     \caption{\label{fig:rosa-wrong}}
 \end{subfigure}
 \caption{(a) The review display. One review has been expanded to display its full text as well as metadata.
 (b) Another review display.~\label{fig:rosa} }
 \end{figure}

Rosa then focused her attention on the reviews listed from this cluster. She wanted to find reviews that mentioned room-quality, so she searched for ``room'' in the search bar and clicked on the first review. She read a useful quote and copies it for her project: \textit{``the rooms aren't huge, granted - but they are clean and well tended.''} (Fig~\ref{fig:rosa-review-1}).

Rosa then wanted to identify some fields where the hotels are under-performing. She looked at the average attribute ratings for each cluster, and noticed that while still positive, cluster 3 had a much lower than average sentiment score for ``location,'' a score of 0.05 vs 0.4 She looked through the reviews themselves, and noticed that several have been flagged with a red ``location'' bubble, which meant that the attribute extractor that she was using had identified that the review expressed a negative opinion about the location of the hotel. However, when she clicked on one review to read it in full, she saw that the review actually had no complaints about the hotel's location, and even said that the reviewer praised the hotel's \textit{``convenient location and good value''} (Fig~\ref{fig:rosa-wrong}). She saw that the review actually complains about the view; Rosa checked her schema and confirmed that ``view'' is listed as an attribute separate from ``location.'' She made a note of this error and sent it to her coworkers working on the extraction algorithm.


Finally, Rosa wanted to define a new schema of attributes relevant to the hotel so that her coworkers could improve the performance of their attribute extractor. She noted that based on the average attribute histogram, ``facility,'' ``food,'' ``general,'' ``location,'' and ``staff'' were important features that have strong sentiments associated with them (whether positive or negative), so she copied them into the new schema field. She also noticed that a lot of reviews discussed public transit options near the hotel, but her current schema did not support this feature. Thus, she added ``public-transit'' to her schema and clicked save, generating a text file that she can send to her coworkers. Armed with these insights, Rosa was ready to help the attribute extraction team build a better algorithm.
 
\begin{figure}[!t]
  \includegraphics[width=0.48\textwidth]{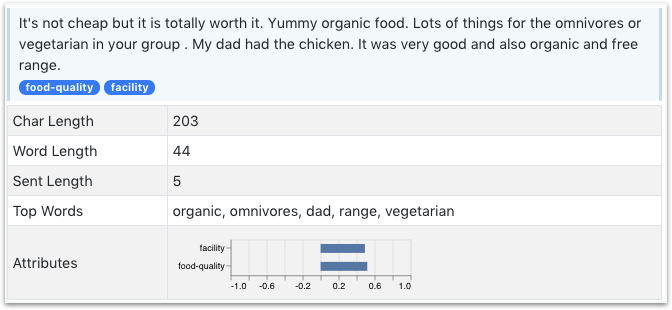}
  \caption{A quick scan of the review section shows that for this restaurant, 
           most of the ``food-quality'' extractions mention vegetarian food, which should 
           be separated to its own category in the schema.\label{fig:vegetarian}}
\end{figure}

\subsection{Designing Extraction Schema for Restaurant Reviews}

We asked Chao to prepare a quick demonstration, using a dataset of restaurant reviews, to demonstrate an attribute extractor and review summarization algorithm that he has been working on. Chao's algorithm already worked for hotel reviews but he needed to design a new schema for the restaurant domain in order to start applying his algorithm on restaurant reviews.

Chao's extractor operates in two stages: discovering aspect-opinion phrase pairs, 
and matching the pairs to categories in a domain-specific schema.
For example, an extracted pair can be (``carpet,'' ``a bit filthy'')
and categorized into the ``cleanliness'' hotel schema attribute.
To apply his extractor to restaurant reviews, he needs to design a new schema, a set of attributes (or features) along with opinion scales, specific to the restaurant domain in order for his extractor to provide useful results. 
First, Chao applied only the first stage of his hotel extractor to the restaurant reviews: extracting aspect-opinion pairs. Based on his knowledge of restaurants, he created a rough attribute schema consisting of ``food,'' ``drink,'' and ``facility,'' and trained a model to categorize the extractions. With the first iteration of his pipeline complete, he loaded these reviews with extractions into Teddy to evaluate the representation power of this schema. The first thing Chao noticed is that one of his review clusters had ``service'' as a commonly occurring word. A grep search for ``service'' confirmed his suspicion that people discuss this aspect frequently, so he added it to his new schema. By clicking on some restaurants and going through a similar discovery process of common terms people use, Chao discovered that his schema needed more granularity: ``food'' was divided into ``food-quality'' and ``healthiness,'' ``drink'' was divided into ``drink'' and ``alcohol,''  and by exploring reviews that mention service he discovered the aspect ``wait-time.''

Next, Chao labeled some examples and trains a classifier for his new schema. Again, he loaded these new extractions into Teddy.

\begin{figure}[!t]
  \includegraphics[width=0.48\textwidth]{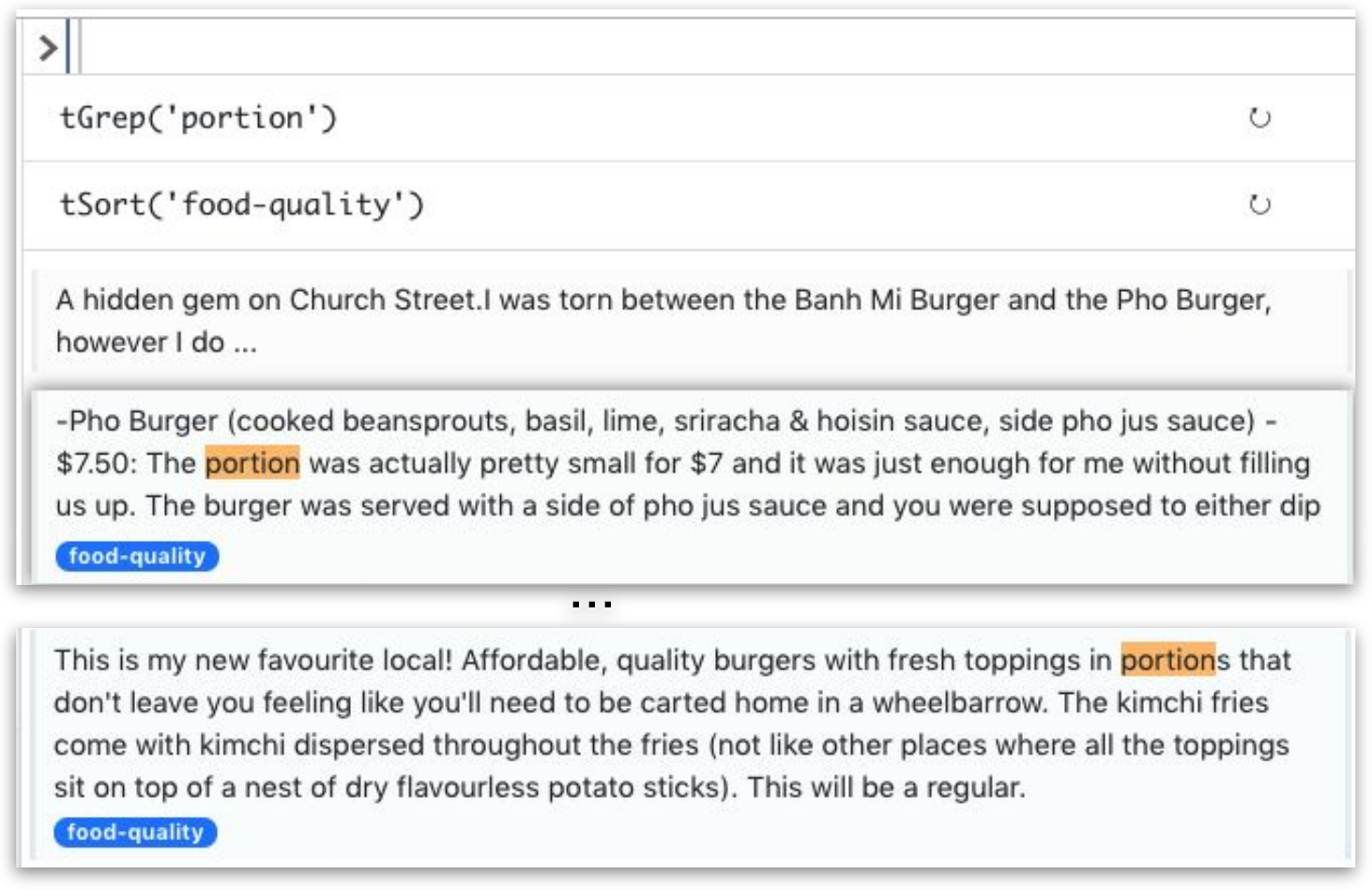}
  \caption{A grep search and sort operation shows that varied opinions 
           in ``food-quality'' can be explained by this newly-discovered 
           aspect relating to portion size. \label{fig:portion}}
\end{figure}

Based on the frequently occurring yet varied terms used for the same valency, Chao decided that ``food-quality'' was still too general. He selected a restaurant that was an outlier with the highest score in its cluster, then sorted the reviews by ``food-quality.'' Several of the reviews mention ``portion size'', and a grep search confirmed that this is a popular aspect, which he further confirmed with another grep search over the full review dataset (Fig~\ref{fig:portion}). Next Chao selected another restaurant with many ``food-quality'' extractions and found that most of the reviews specifically mention plentiful vegetarian options (Fig~\ref{fig:vegetarian}).  He additionally decided on a ``location'' aspect, aided by the map feature which helped him discover that this topic was mentioned more in certain neighborhoods.

Chao re-trained his extraction classifier and evaluated the results on his test set.  
Table~\ref{tab:classifier} shows the accuracies of the restaurant attribute classifiers
when trained on three schema versions with increasing numbers of attributes. Each classifier is trained on 5,000 aspect-opinion pairs
by fine-tuning the uncased 12-layer BERT model~\cite{devlin2019bert} for 10 epochs
and is evaluated on 1,000 labeled pairs.
The three versions consist of 3, 7, and 10 attributes created as described above.
Each schema also contains a special attribute ``others'' indicating examples that are not covered.
The coverage(\%) column shows the percentage of test examples that are not labeled as ``others.''
From increasing the number of attributes from 3 to 10, the coverage increased by 14.2\%.
The last 3 columns show the classifiers' accuracy.
Each column ``Level-$i$'' indicates the accuracy when evaluated at the granularity of schema-$i$.
When a classifier is evaluated at a less fine-grained (easier) level,
each predicted label is mapped to the attribute that contains the predicted label (e.g., ``food-quality'' to ``food,'' or ``alcohol'' to ``drink'').
When evaluated at a finer-grained (harder) level than the classifier's schema,
each predicted label is mapped to a single attribute 
(For example, when schema-1 is evaluated on Level-3, a ``drink'' or ``facility'' prediction is only correct if those are the ground-truth labels, but a ``food'' prediction is only correct if the ground-truth is ``food-quality,'' since ``food'' is not in the level-3 schema).
Chao's results clearly showed that as the schema became more carefully designed, the accuracy of the BERT classifier increased significantly (by 12.9\% at Level-2 and by 16.9\% at Level-3).

\setlength{\tabcolsep}{3.5pt}
\begin{table}[!t]
\centering
\begin{tabular}{cccccc}\toprule
& \#attr   & coverage(\%) & Level-1 & Level-2 & Level-3 \\ \midrule
Schema-1 & 3        & 78.4  & 90.9  & 72.6  & 68.0 \\
Schema-2 & 7        & 92.6  & 88.9  & 84.5  & 80.3 \\
Schema-3 & 10       & \textbf{92.6}  & 90.6  & \textbf{85.5}  & \textbf{84.9} \\ \bottomrule
\end{tabular}
\caption{BERT classifiers' accuracy using different schemas.}
\label{tab:classifier}
\end{table}

Loading these new extractions into Teddy, he noticed more granular differences between clusters. For example, the two largest clusters were already clearly separated by average sentiment, but now Chao observed that much of the difference in sentiment could be attributed to the aspect ``service'' (Fig~\ref{fig:food_clusters}). With his new schema, Chao was able to prepare a better extraction and summarization demonstration for a prospective client.

%% file: discussion.tex
\section{Discussion and Conclusion\label{sec:discussion}}

The size and availability of user-generated reviews on the Web are rapidly increasing with the expansion of e-commerce applications. Combined with the wealth and the potential utility of information in reviews, this trend has created a widespread demand in review text analysis and mining.

In this paper, we first contribute findings from an interview study  conducted with fifteen data scientists to better understand the current  review analysis practices and challenges. To the best of our knowledge, the study presented here is the first interview study focused on review text analysis and mining. 

Our results suggest that data scientists performing review analyses are not very concerned developing new models, architectures, or parameter tuning and that they are largely satisfied with using existing language models. They often operate on the assumption that improving the quality of their training data is the most effective means to improve the performance of their models. On the other hand, they are hampered by the lack of tools that would help them across different stages of data preparation.

Also, perhaps unsurprisingly, our study led us to the conclusion that it is very difficult to build a single tool that fits all data scientists' needs.  Challenges and priorities vary across the roles and analysis goals of data scientists.  For example, pipeline management and provenance is an important concern motivated by the cost of context-switching among senior/lead  data scientists. 
On the other hand, junior data scientists are more concerned about data labeling and crowdsourcing as bottlenecks than senior data scientists are. 

Results also indicate that data scientists spend most of their time on data preparation, lending additional support for the currently accepted general wisdom and extending it to the special case of review text analysis.  However, our results also contribute further,  finer-grained insights on data preparation for the review domain.  For example, we find that the most time consuming or challenging data preparation steps are labeling, designing and specifying crowdsourcing tasks, and interactive exploration, not necessarily cleaning.

Through our interview study, we also find that data scientists lack interactive tools with which to quickly explore large collections of reviews  together with results of extractors and models on these reviews. In response, we contribute Teddy, an interactive system to help data scientists gain 
insights into review text at scale along with fine-grained opinions expressed in reviews and enable to data scientists iteratively  refine and improve their extraction schemas and models. We demonstrate the utility of Teddy in depth through two use cases carried out by data scientists.  Informed by the interview study results, we believe that Teddy addresses an important need in review analysis pipelines. 

Products or services purchased on the Web are increasingly defined by their reviews. Reviews (or any other user-generated text referring to user experiences with entities, for that matter) are becoming 
distributional representations of these products or services, analogous to distributional semantics~\cite{firth1957synopsis}. Research efforts into better tools for understanding and mining the rich information embodied in user generated 
text, such as Teddy, are therefore essential to the future of improved user experience on the Web. To support extended research and applications, we make Teddy and the data collected from our interview study publicly available at~\url{https://github.com/megagonlabs/teddy}. 

\begin{figure}[!t]
\centering 
\includegraphics[width=0.48\textwidth]{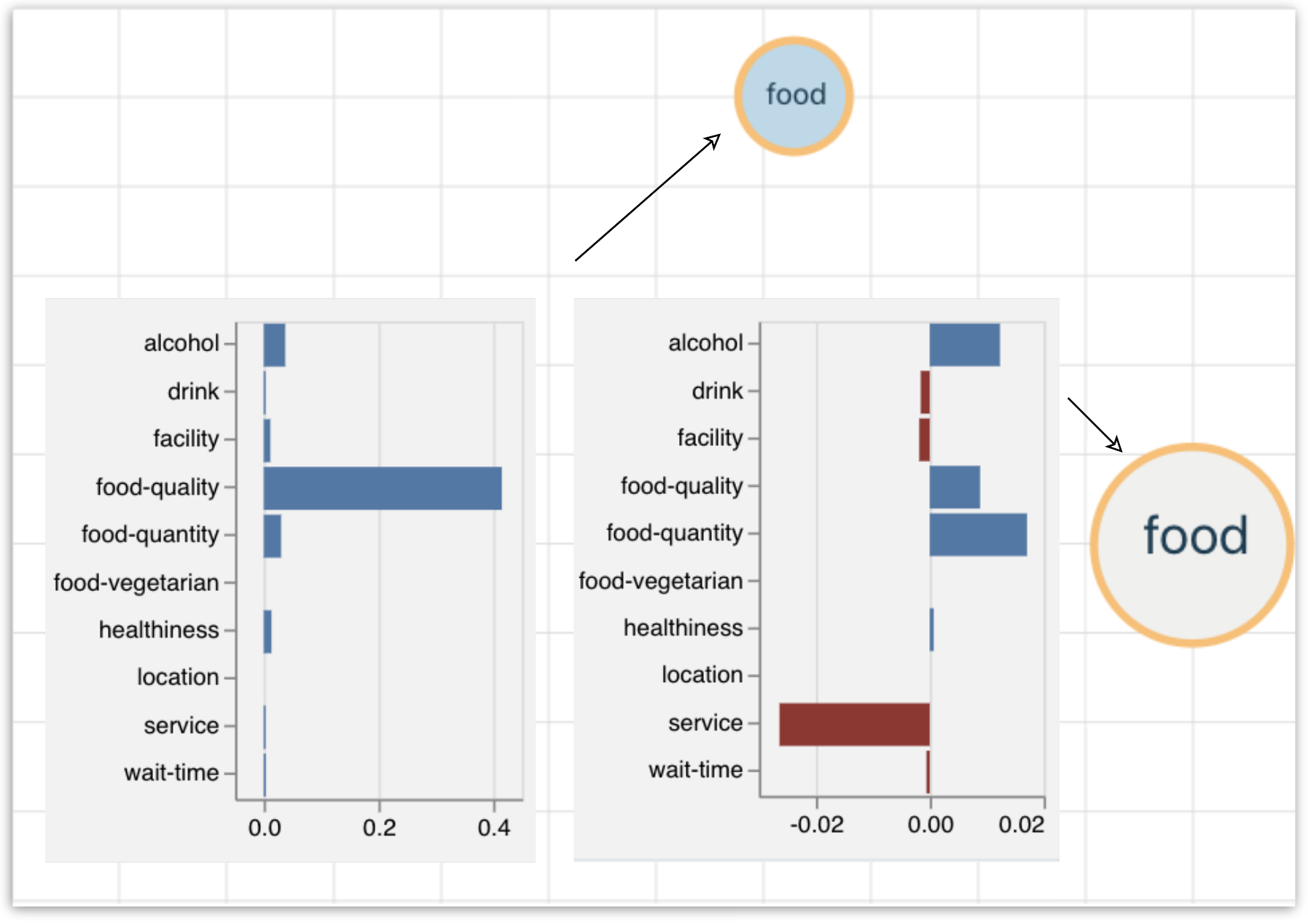}
\caption{The newly-designed schema reveals the differences between the two 
largest clusters, which were unclear in previous iterations but are now 
largely explained by differences in service.\label{fig:food_clusters}}
\end{figure}

%% file: ack.tex
\section{Acknowledgments\label{sec:ack}} 
We thank our study participants for their time and insights. We also thank Eser Kando\u{g}an for his feedback on an earlier draft of this paper. 

%% file: main.bbl

\begin{thebibliography}{00}


\ifx \showCODEN    \undefined \def \showCODEN     #1{\unskip}     \fi
\ifx \showDOI      \undefined \def \showDOI       #1{{\tt DOI:}\penalty0{#1}\ }
  \fi
\ifx \showISBNx    \undefined \def \showISBNx     #1{\unskip}     \fi
\ifx \showISBNxiii \undefined \def \showISBNxiii  #1{\unskip}     \fi
\ifx \showISSN     \undefined \def \showISSN      #1{\unskip}     \fi
\ifx \showLCCN     \undefined \def \showLCCN      #1{\unskip}     \fi
\ifx \shownote     \undefined \def \shownote      #1{#1}          \fi
\ifx \showarticletitle \undefined \def \showarticletitle #1{#1}   \fi
\ifx \showURL      \undefined \def \showURL       #1{#1}          \fi

\bibitem{d3}
 2019.
\newblock Data-Driven Documents.
\newblock \url{https://d3js.org/}.   (2019).
\newblock
\newblock
\shownote{Accessed: 2019-12-25.}


\bibitem{flask}
 2019.
\newblock Flask - Full Stack Python.
\newblock \url{https://www.fullstackpython.com/flask.html}.   (2019).
\newblock
\newblock
\shownote{Accessed: 2019-12-25.}


\bibitem{jupyter}
 2019.
\newblock Project Jupyter.
\newblock \url{http://jupyter.org/}.   (2019).
\newblock
\newblock
\shownote{Accessed: 2019-12-25.}


\bibitem{pandas}
 2019.
\newblock Python Data Analysis Library.
\newblock \url{https://pandas.pydata.org/}.   (2019).
\newblock
\newblock
\shownote{Accessed: 2019-12-25.}


\bibitem{react}
 2019.
\newblock React - A JavaScript library for building user interfaces.
\newblock \url{https://reactjs.org/}.   (2019).
\newblock
\newblock
\shownote{Accessed: 2019-12-25.}


\bibitem{scikitlearn}
 2019.
\newblock scikit-learn machine learning in Python.
\newblock \url{https://scikit-learn.org/stable/}.   (2019).
\newblock
\newblock
\shownote{Accessed: 2019-12-25.}


\bibitem{vegalite}
 2019.
\newblock Vega-Lite A Grammar of Interactive Graphics.
\newblock \url{https://vega.github.io/vega-lite/}.   (2019).
\newblock
\newblock
\shownote{Accessed: 2019-12-25.}


\bibitem{alper2011opinionblocks}
{Basak Alper}, {Huahai Yang}, {Eben Haber}, {and} {Eser Kandogan}. 2011.
\newblock \showarticletitle{OpinionBlocks: Visualizing consumer reviews}. In
  {\em IEEE VisWeek Workshop on Interactive Visual Text Analytics for Decision
  Making}.
\newblock


\bibitem{BirdKleinLoper09}
{Steven Bird}, {Ewan Klein}, {and} {Edward Loper}. 2009.
\newblock {\em Natural Language Processing with Python\/} (1st ed.).
\newblock O'Reilly Media, Inc.
\newblock
\showISBNx{0596516495, 9780596516499}


\bibitem{blei2003latent}
{David~M. Blei}, {Andrew~Y. Ng}, {and} {Michael~I. Jordan}. 2003.
\newblock \showarticletitle{Latent Dirichlet Allocation}.
\newblock {\em J. Mach. Learn. Res.\/}  {3} (2003), 993--1022.
\newblock
\showURL{%
\url{http://jmlr.org/papers/v3/blei03a.html}}


\bibitem{di2013sentiment}
{Luigi~Di Caro} {and} {Matteo Grella}. 2013.
\newblock \showarticletitle{Sentiment analysis via dependency parsing}.
\newblock {\em Computer Standards {\&} Interfaces\/} {35}, 5 (2013), 442--453.
\newblock
\showDOI{%
\url{http://dx.doi.org/10.1016/j.csi.2012.10.005}}


\bibitem{chen2006visual}
{Chaomei Chen}, {Fidelia Ibekwe{-}Sanjuan}, {Eric SanJuan}, {and} {Chris
  Weaver}. 2006.
\newblock \showarticletitle{Visual Analysis of Conflicting Opinions}. In {\em
  {IEEE} Symposium On Visual Analytics Science And Technology, {IEEE} {VAST}
  2006, October 31-November 2, 2006, Baltimore, Maryland, {USA}}. 59--66.
\newblock
\showDOI{%
\url{http://dx.doi.org/10.1109/VAST.2006.261431}}


\bibitem{chuang2012interpretation}
{Jason Chuang}, {Daniel Ramage}, {Christopher Manning}, {and} {Jeffrey Heer}.
  2012.
\newblock \showarticletitle{Interpretation and trust: Designing model-driven
  visualizations for text analysis}. In {\em Proceedings of the SIGCHI
  Conference on Human Factors in Computing Systems}. ACM, 443--452.
\newblock


\bibitem{collins2009docuburst}
{Christopher Collins}, {Sheelagh Carpendale}, {and} {Gerald Penn}. 2009a.
\newblock \showarticletitle{Docuburst: Visualizing document content using
  language structure}. In {\em Computer graphics forum}, Vol.~28. Wiley Online
  Library, 1039--1046.
\newblock


\bibitem{collins2009parallel}
{Christopher Collins}, {Fernanda~B Viegas}, {and} {Martin Wattenberg}. 2009b.
\newblock \showarticletitle{Parallel tag clouds to explore and analyze faceted
  text corpora}. In {\em 2009 IEEE Symposium on Visual Analytics Science and
  Technology}. IEEE, 91--98.
\newblock


\bibitem{crouch2006}
{Mira Crouch} {and} {Heather McKenzie}. 2006.
\newblock \showarticletitle{The logic of small samples in interview-based
  qualitative research}.
\newblock {\em Social Science Information\/} {45}, 4 (2006), 483--499.
\newblock


\bibitem{Demiralp:2017:DSIA}
{{\c{C}}agatay Demiralp}, {Peter~J. Haas}, {Srinivasan Parthasarathy}, {and}
  {Tejaswini Pedapati}. 2017.
\newblock \showarticletitle{Foresight: Rapid Data Exploration Through
  Guideposts}.
\newblock {\em CoRR\/}  {abs/1709.10513} (2017).
\newblock
\showURL{%
\url{http://arxiv.org/abs/1709.10513}}


\bibitem{devlin2019bert}
{Jacob Devlin}, {Ming{-}Wei Chang}, {Kenton Lee}, {and} {Kristina Toutanova}.
  2019.
\newblock \showarticletitle{{BERT:} Pre-training of Deep Bidirectional
  Transformers for Language Understanding}. In {\em Proceedings of the 2019
  Conference of the North American Chapter of the Association for Computational
  Linguistics: Human Language Technologies, {NAACL-HLT} 2019, Minneapolis, MN,
  USA, June 2-7, 2019, Volume 1 (Long and Short Papers)}. 4171--4186.
\newblock
\showURL{%
\url{https://www.aclweb.org/anthology/N19-1423/}}


\bibitem{dou2013hierarchicaltopics}
{Wenwen Dou}, {Li Yu}, {Xiaoyu Wang}, {Zhiqiang Ma}, {and} {William Ribarsky}.
  2013.
\newblock \showarticletitle{HierarchicalTopics: Visually exploring large text
  collections using topic hierarchies}.
\newblock {\em IEEE Transactions on Visualization and Computer Graphics\/}
  {19}, 12 (2013), 2002--2011.
\newblock


\bibitem{felix2016texttile}
{Cristian Felix}, {Anshul~Vikram Pandey}, {and} {Enrico Bertini}. 2017.
\newblock \showarticletitle{TextTile: An Interactive Visualization Tool for
  Seamless Exploratory Analysis of Structured Data and Unstructured Text}.
\newblock {\em {IEEE} Trans. Vis. Comput. Graph.\/} {23}, 1 (2017), 161--170.
\newblock
\showDOI{%
\url{http://dx.doi.org/10.1109/TVCG.2016.2598447}}


\bibitem{firth1957synopsis}
{John~R Firth}. 1957.
\newblock \showarticletitle{A synopsis of linguistic theory, 1930-1955}.
\newblock {\em Studies in linguistic analysis\/} (1957).
\newblock


\bibitem{gamon2005pulse}
{Michael Gamon}, {Anthony Aue}, {Simon Corston{-}Oliver}, {and} {Eric~K.
  Ringger}. 2005.
\newblock \showarticletitle{Pulse: Mining Customer Opinions from Free Text}. In
  {\em Advances in Intelligent Data Analysis VI, 6th International Symposium on
  Intelligent Data Analysis, {IDA} 2005, Madrid, Spain, September 8-10, 2005,
  Proceedings}. 121--132.
\newblock
\showDOI{%
\url{http://dx.doi.org/10.1007/11552253\_12}}


\bibitem{gregory2006user}
{Michelle~L. Gregory}, {Nancy Chinchor}, {Paul Whitney}, {Richard Carter},
  {Elizabeth Hetzler}, {and} {Alan Turner}. 2006.
\newblock \showarticletitle{User-directed Sentiment Analysis: Visualizing the
  Affective Content of Documents}. In {\em Proceedings of the Workshop on
  Sentiment and Subjectivity in Text} {\em (SST '06)}. Association for
  Computational Linguistics, Stroudsburg, PA, USA, 23--30.
\newblock
\showISBNx{1-932432-75-2}
\showURL{%
\url{http://dl.acm.org/citation.cfm?id=1654641.1654645}}


\bibitem{guest2006}
{Bunce~A. Guest, G.} {and} {L. Johnson}. 2006.
\newblock \showarticletitle{How many interviews are enough? An experiment with
  data saturation and variability}.
\newblock {\em Field Methods\/} {18}, 1 (2006), 59--82.
\newblock


\bibitem{hao2013visual}
{Ming~C. Hao}, {Christian Rohrdantz}, {Halldor Janetzko}, {Daniel~A. Keim},
  {Umeshwar Dayal}, {Lars{-}Erik Haug}, {Meichun Hsu}, {and} {Florian Stoffel}.
  2013.
\newblock \showarticletitle{Visual sentiment analysis of customer feedback
  streams using geo-temporal term associations}.
\newblock {\em Information Visualization\/} {12}, 3-4 (2013), 273--290.
\newblock
\showDOI{%
\url{http://dx.doi.org/10.1177/1473871613481691}}


\bibitem{havre2000themeriver}
{Susan Havre}, {Beth Hetzler}, {and} {Lucy Nowell}. 2000.
\newblock \showarticletitle{ThemeRiver: Visualizing theme changes over time}.
  In {\em IEEE Symposium on Information Visualization 2000. INFOVIS 2000.
  Proceedings}. IEEE, 115--123.
\newblock


\bibitem{spacy2019}
{Matthew Honnibal}. 2015.
\newblock spaCy: Industrial-strength Natural Language Processing (NLP) with
  Python and Cython.
\newblock   (2015).
\newblock


\bibitem{hu2004mining}
{Minqing Hu} {and} {Bing Liu}. 2004a.
\newblock \showarticletitle{Mining and summarizing customer reviews}. In {\em
  Proceedings of the Tenth {ACM} {SIGKDD} International Conference on Knowledge
  Discovery and Data Mining, Seattle, Washington, USA, August 22-25, 2004}.
  168--177.
\newblock
\showDOI{%
\url{http://dx.doi.org/10.1145/1014052.1014073}}


\bibitem{hu2004features}
{Minqing Hu} {and} {Bing Liu}. 2004b.
\newblock \showarticletitle{Mining Opinion Features in Customer Reviews}. In
  {\em Proceedings of the Nineteenth National Conference on Artificial
  Intelligence, Sixteenth Conference on Innovative Applications of Artificial
  Intelligence, July 25-29, 2004, San Jose, California, {USA}}. 755--760.
\newblock
\showURL{%
\url{http://www.aaai.org/Library/AAAI/2004/aaai04-119.php}}


\bibitem{kandel2012enterprise}
{Sean Kandel}, {Andreas Paepcke}, {Joseph~M. Hellerstein}, {and} {Jeffrey
  Heer}. 2012.
\newblock \showarticletitle{Enterprise Data Analysis and Visualization: An
  Interview Study}.
\newblock {\em {IEEE} Trans. Vis. Comput. Graph.\/} {18}, 12 (2012),
  2917--2926.
\newblock
\showDOI{%
\url{http://dx.doi.org/10.1109/TVCG.2012.219}}


\bibitem{kandogan2014data}
{Eser Kandogan}, {Aruna Balakrishnan}, {Eben~M. Haber}, {and} {Jeffrey~S.
  Pierce}. 2014.
\newblock \showarticletitle{From Data to Insight: Work Practices of Analysts in
  the Enterprise}.
\newblock {\em {IEEE} Computer Graphics and Applications\/} {34}, 5 (2014),
  42--50.
\newblock
\showDOI{%
\url{http://dx.doi.org/10.1109/MCG.2014.62}}


\bibitem{kim2004determining}
{Soo{-}Min Kim} {and} {Eduard~H. Hovy}. 2004.
\newblock \showarticletitle{Determining the Sentiment of Opinions}. In {\em
  {COLING} 2004, 20th International Conference on Computational Linguistics,
  Proceedings of the Conference, 23-27 August 2004, Geneva, Switzerland}.
\newblock
\showURL{%
\url{https://www.aclweb.org/anthology/C04-1200/}}


\bibitem{kucher2018state}
{Kostiantyn Kucher}, {Carita Paradis}, {and} {Andreas Kerren}. 2018.
\newblock \showarticletitle{The State of the Art in Sentiment Visualization}.
\newblock {\em Comput. Graph. Forum\/} {37}, 1 (2018), 71--96.
\newblock
\showDOI{%
\url{http://dx.doi.org/10.1111/cgf.13217}}


\bibitem{li2019subjective}
{Yuliang Li}, {Aaron Feng}, {Jinfeng Li}, {Saran Mumick}, {Alon~Y. Halevy},
  {Vivian Li}, {and} {Wang{-}Chiew Tan}. 2019.
\newblock \showarticletitle{Subjective Databases}.
\newblock {\em {PVLDB}\/} {12}, 11 (2019), 1330--1343.
\newblock
\showDOI{%
\url{http://dx.doi.org/10.14778/3342263.3342271}}


\bibitem{liu2005opinion}
{Bing Liu}, {Minqing Hu}, {and} {Junsheng Cheng}. 2005.
\newblock \showarticletitle{Opinion observer: analyzing and comparing opinions
  on the Web}. In {\em Proceedings of the 14th international conference on
  World Wide Web, {WWW} 2005, Chiba, Japan, May 10-14, 2005}. 342--351.
\newblock
\showDOI{%
\url{http://dx.doi.org/10.1145/1060745.1060797}}


\bibitem{liu2012survey}
{Bing Liu} {and} {Lei Zhang}. 2012.
\newblock \showarticletitle{A Survey of Opinion Mining and Sentiment Analysis}.
\newblock In {\em Mining Text Data}. 415--463.
\newblock
\showDOI{%
\url{http://dx.doi.org/10.1007/978-1-4614-3223-4\_13}}


\bibitem{liu2018bridging}
{Shixia Liu}, {Xiting Wang}, {Christopher Collins}, {Wenwen Dou}, {Fang{-}Xin
  Ou{-}Yang}, {Mennatallah El{-}Assady}, {Liu Jiang}, {and} {Daniel~A. Keim}.
  2019.
\newblock \showarticletitle{Bridging Text Visualization and Mining: {A}
  Task-Driven Survey}.
\newblock {\em {IEEE} Trans. Vis. Comput. Graph.\/} {25}, 7 (2019), 2482--2504.
\newblock
\showDOI{%
\url{http://dx.doi.org/10.1109/TVCG.2018.2834341}}


\bibitem{morinaga2002mining}
{Satoshi Morinaga}, {Kenji Yamanishi}, {Kenji Tateishi}, {and} {Toshikazu
  Fukushima}. 2002.
\newblock \showarticletitle{Mining product reputations on the Web}. In {\em
  Proceedings of the Eighth {ACM} {SIGKDD} International Conference on
  Knowledge Discovery and Data Mining, July 23-26, 2002, Edmonton, Alberta,
  Canada}. 341--349.
\newblock
\showDOI{%
\url{http://dx.doi.org/10.1145/775047.775098}}


\bibitem{muller2019data}
{Michael~J. Muller}, {Ingrid Lange}, {Dakuo Wang}, {David Piorkowski}, {Jason
  Tsay}, {Q.~Vera Liao}, {Casey Dugan}, {and} {Thomas Erickson}. 2019.
\newblock \showarticletitle{How Data Science Workers Work with Data: Discovery,
  Capture, Curation, Design, Creation}. In {\em Proceedings of the 2019 {CHI}
  Conference on Human Factors in Computing Systems, {CHI} 2019, Glasgow,
  Scotland, UK, May 04-09, 2019}. 126.
\newblock
\showDOI{%
\url{http://dx.doi.org/10.1145/3290605.3300356}}


\bibitem{oelke2009visual}
{Daniela Oelke}, {Ming~C. Hao}, {Christian Rohrdantz}, {Daniel~A. Keim},
  {Umeshwar Dayal}, {Lars{-}Erik Haug}, {and} {Halld{\'{o}}r Janetzko}. 2009.
\newblock \showarticletitle{Visual opinion analysis of customer feedback data}.
  In {\em Proceedings of the {IEEE} Symposium on Visual Analytics Science and
  Technology, {IEEE} {VAST} 2009, Atlantic City, New Jersey, USA, 11-16 October
  2009, part of VisWeek 2009}. 187--194.
\newblock
\showDOI{%
\url{http://dx.doi.org/10.1109/VAST.2009.5333919}}


\bibitem{pang2008opinion}
{Bo Pang} {and} {Lillian Lee}. 2008.
\newblock \showarticletitle{Opinion Mining and Sentiment Analysis}.
\newblock {\em Foundations and Trends in Information Retrieval\/}  {2} (2008),
  1--135.
\newblock


\bibitem{pang2002thumbs}
{Bo Pang}, {Lillian Lee}, {and} {Shivakumar Vaithyanathan}. 2002.
\newblock \showarticletitle{Thumbs up? Sentiment Classification using Machine
  Learning Techniques}. In {\em Proceedings of the 2002 Conference on Empirical
  Methods in Natural Language Processing, {EMNLP} 2002, Philadelphia, PA, USA,
  July 6-7, 2002}.
\newblock
\showURL{%
\url{https://www.aclweb.org/anthology/W02-1011/}}


\bibitem{popescu2005extracting}
{Ana{-}Maria Popescu} {and} {Oren Etzioni}. 2005.
\newblock \showarticletitle{Extracting Product Features and Opinions from
  Reviews}. In {\em {HLT/EMNLP} 2005, Human Language Technology Conference and
  Conference on Empirical Methods in Natural Language Processing, Proceedings
  of the Conference, 6-8 October 2005, Vancouver, British Columbia, Canada}.
  339--346.
\newblock
\showURL{%
\url{https://www.aclweb.org/anthology/H05-1043/}}


\bibitem{rehurek_lrec}
{Radim {\v R}eh{\r u}{\v r}ek} {and} {Petr Sojka}. 2010.
\newblock \showarticletitle{{Software Framework for Topic Modelling with Large
  Corpora}}. In {\em {Proceedings of the LREC 2010 Workshop on New Challenges
  for NLP Frameworks}}. ELRA, Valletta, Malta, 45--50.
\newblock


\bibitem{Shneiderman:1996:ETD}
{Ben Shneiderman}. 1996.
\newblock \showarticletitle{The Eyes Have It: {A} Task by Data Type Taxonomy
  for Information Visualizations}. In {\em Proceedings of the 1996 {IEEE}
  Symposium on Visual Languages, Boulder, Colorado, USA, September 3-6, 1996}.
  336--343.
\newblock
\showDOI{%
\url{http://dx.doi.org/10.1109/VL.1996.545307}}


\bibitem{soto2015exploratory}
{Axel~J. Soto}, {Ryan Kiros}, {Vlado Keselj}, {and} {Evangelos~E. Milios}.
  2015.
\newblock \showarticletitle{Exploratory Visual Analysis and Interactive Pattern
  Extraction from Semi-Structured Data}.
\newblock {\em TiiS\/} {5}, 3 (2015), 16:1--16:36.
\newblock
\showDOI{%
\url{http://dx.doi.org/10.1145/2812115}}


\bibitem{stasko2008jigsaw}
{John Stasko}, {Carsten G{\"o}rg}, {and} {Zhicheng Liu}. 2008a.
\newblock \showarticletitle{Jigsaw: supporting investigative analysis through
  interactive visualization}.
\newblock {\em Information visualization\/} {7}, 2 (2008), 118--132.
\newblock


\bibitem{stasko2008textsense}
{John Stasko}, {Carsten G{\"o}rg}, {and} {Zhicheng Liu}. 2008b.
\newblock \showarticletitle{Sensemaking across text documents: human-centered,
  visual exploration with Jigsaw}. In {\em CHI'08 Workshop on Sensemaking}.
\newblock


\bibitem{TAKAOKA18.8884}
{Kazuma Takaoka}, {Sorami Hisamoto}, {Noriko Kawahara}, {Miho Sakamoto},
  {Yoshitaka Uchida}, {and} {Yuji Matsumoto}. 2018.
\newblock \showarticletitle{Sudachi: a Japanese Tokenizer for Business}. In
  {\em Proceedings of the Eleventh International Conference on Language
  Resources and Evaluation, {LREC} 2018, Miyazaki, Japan, May 7-12, 2018}.
\newblock
\showURL{%
\url{http://www.lrec-conf.org/proceedings/lrec2018/summaries/8884.html}}


\bibitem{wattenberg2008word}
{Martin Wattenberg} {and} {Fernanda~B Vi{\'e}gas}. 2008.
\newblock \showarticletitle{The word tree, an interactive visual concordance}.
\newblock {\em IEEE transactions on visualization and computer graphics\/}
  {14}, 6 (2008), 1221--1228.
\newblock


\bibitem{wu2010opinionseer}
{Yingcai Wu}, {Furu Wei}, {Shixia Liu}, {Norman Au}, {Weiwei Cui}, {Hong Zhou},
  {and} {Huamin Qu}. 2010.
\newblock \showarticletitle{OpinionSeer: Interactive Visualization of Hotel
  Customer Feedback}.
\newblock {\em {IEEE} Trans. Vis. Comput. Graph.\/} {16}, 6 (2010), 1109--1118.
\newblock
\showDOI{%
\url{http://dx.doi.org/10.1109/TVCG.2010.183}}


\bibitem{yatani2011review}
{Koji Yatani}, {Michael Novati}, {Andrew Trusty}, {and} {Khai~N. Truong}. 2011.
\newblock \showarticletitle{Review Spotlight: a user interface for summarizing
  user-generated reviews using adjective-noun word pairs}. In {\em Proceedings
  of the International Conference on Human Factors in Computing Systems, {CHI}
  2011, Vancouver, BC, Canada, May 7-12, 2011}. 1541--1550.
\newblock
\showDOI{%
\url{http://dx.doi.org/10.1145/1978942.1979167}}


\end{thebibliography}
